%% file: arXiv.tex
\documentclass[12pt]{article}

\RequirePackage{etex}

\usepackage[utf8]{inputenc}
\usepackage[T1]{fontenc}

\usepackage[margin=1in]{geometry}
\usepackage{parskip}

\usepackage{textcomp}

\usepackage[title]{appendix}

\usepackage[scaled=0.92]{helvet}

\usepackage{mathtools}
\usepackage{bm}
\usepackage{pdflscape}

\usepackage{xfrac}
\usepackage{pdfpages}
\usepackage{siunitx}
\usepackage{enumitem}
\usepackage[export]{adjustbox}
\usepackage{needspace}
\usepackage{tabularx}
\usepackage{wrapfig}
\usepackage{lscape}
\usepackage[referable]{threeparttablex}
\usepackage{titlesec}
\usepackage{setspace}
\usepackage{longtable}
\usepackage{multirow}
\usepackage{booktabs}
\usepackage{afterpage}
\usepackage{relsize}
\usepackage{cprotect}
\usepackage{array}
\usepackage{fancyhdr}
\usepackage{lipsum}
\usepackage{subcaption}
\usepackage[labelfont=bf]{caption}

\usepackage[numbers,sort&compress]{natbib}
\usepackage[
    colorlinks=true,
    linkcolor=red,
    citecolor=blue,
    urlcolor=blue
]{hyperref}

\titleformat{\section}
  {\normalfont\Large\bfseries}{\thesection}{1em}{}

\newcolumntype{P}[1]{>{\centering\arraybackslash}p{#1}}

\allowdisplaybreaks
\fancypagestyle{landscape}{
  \fancyhf{}
  \fancyfoot[C]{\thepage}
  \renewcommand{\headrulewidth}{0pt}
  
}

\usepackage{autonum}

\title{\bf A flexible quantile mixed-effects model for censored outcomes}

\author{
  Divan A.\ Burger\textsuperscript{*\textdagger} \and
  Sean van der Merwe\textsuperscript{\textdagger} \and
  Emmanuel Lesaffre\textsuperscript{\textdaggerdbl}
}

\date{}

\providecommand{\bibcommenthead}{}

\begin{document}

    \onehalfspacing
    
    \maketitle

    {\footnotesize
        \noindent\textsuperscript{*} Syneos Health, Bloemfontein, Free State, South Africa \\
        \textsuperscript{\textdagger} Department of Mathematical Statistics and Actuarial Science, 
        University of the Free State, Bloemfontein, Free State, South Africa \\
        \textsuperscript{\textdaggerdbl} I-BioStat, KU Leuven, Leuven, Belgium; 
        Department of Statistics and Actuarial Science, University of Stellenbosch, Stellenbosch, South Africa
    
        \noindent\textbf{Corresponding author:} Divan Burger, Syneos Health, Bloemfontein, South Africa 9301. \\ \textbf{Email:} \texttt{divanaburger@gmail.com} \\
    }

    \begin{abstract}
        \noindent
        We introduce a Bayesian quantile mixed-effects model for censored longitudinal outcomes based on the skew exponential power (SEP) error distribution. The SEP family separates tail behavior and skewness from the targeted quantile and includes the skew Laplace (SL) distribution as a special case. We derive analytic likelihood contributions for left, right, and interval censoring under the SEP model, so censored observations are handled within a single parametric framework without numerical integration in the likelihood. In simulation studies with varying censoring patterns and skewness profiles, the SEP-based quantile mixed-effects model maintains near-nominal bias and credible interval coverage for regression coefficients. In contrast, the SL-based model can exhibit bias and undercoverage when the data’s skewness conflicts with the skewness implied by the target quantile. In an HIV-1 RNA viral load case study with left censoring at the assay limit, bridge-sampled marginal likelihoods and simulation-based residual diagnostics favor the SEP specification across quantiles and yield more stable estimates of treatment-specific viral load trajectories than the SL benchmark.
    \end{abstract}


    \section{Introduction}

    Quantile regression has become a standard tool for examining how covariate effects differ across the entire conditional distribution of a response rather than being summarized by a single mean effect \citep{KOENKER1978, koenker_2005}. In longitudinal and other hierarchical settings, mixed-effects extensions enable researchers to separate within- and between-subject variation, making them especially useful whenever measurements are collected over time.

    Many quantile mixed-effects models (QMMs) adopt the skew Laplace (SL) distribution (also known as the asymmetric Laplace) \citep{GERACI2007}. The SL is attractive because its location parameter $\mu$ coincides with the target quantile $p_{0}$. However, that convenience carries a cost: once $p_{0}$ is fixed, the distribution's skewness is also fixed. For instance, setting $p_{0}=0.50$ yields a symmetric Laplace density, whereas choosing an upper quantile ($p_{0}>0.50$) forces left-skewness and a lower quantile ($p_{0}<0.50$) forces right-skewness. If the data show a different pattern, this built-in link can lead to a misfit, and the SL's moderately heavy tails give limited protection against extreme values.
    
    Censoring in the application of QMMs remains a hurdle. Fully parametric QMMs that handle left, right, or interval censoring currently exist only under the SL family. Representative examples are the longitudinal model of Galarza et al. \citep{galarza2020quantile}, the shared frailty extension of Yazdani et al. \citep{yazdani2022laplace}, and the joint longitudinal-survival formulation of Zhang and Huang \citep{zhang2021bayesian}. Other work, such as the clustered failure-time method of Yin and Cai \citep{yin2005quantile}, the recurrent gap-time model of Luo et al. \citep{luo2013quantile}, and the measurement error approach of Tian et al. \citep{tian2018joint}, relies on inverse-probability weighting or related estimating-equation techniques. Because these semiparametric methods lack a full likelihood with an analytic CDF, they cannot provide closed-form censored likelihood terms or marginal likelihoods for model comparison; hence, no fully parametric censored QMM exists outside the SL family.

    Several authors have attempted to relax the SL tails, for example, the generalized SL of Yan et al. \citep{yan2025new} and its contaminated version \citep{burger2025robust}. These variants, however, still keep skewness largely tied to the chosen quantile and have been developed only for fixed-effects models.

    The skew exponential power (SEP) distribution \citep{zhu2009properties} resolves both problems: it separates skewness and tail thickness from the target quantile through two shape parameters, and it provides a closed-form CDF that makes censored likelihood contributions analytic. In this paper, we extend the SEP distribution to a fully parametric mixed-effects framework that handles censoring; previous studies employed the SEP only in uncensored fixed-effects quantile models \citep{bernardi2018bayesian, arnroth2023quantile}.

    We address the unmet need for a fully parametric QMM that both employs the SEP error distribution and handles left-, right-, and interval-censoring analytically by developing a Bayesian SEP-based framework implemented in \texttt{JAGS} with weakly informative priors.

    We benchmark the new model against the SL special case on longitudinal HIV viral load data, where $\log_{10}$ RNA copies decline biphasically, and many observations are left-censored. CD4 counts enter as time-varying covariates. We use a simulation study to show that the SEP QMM maintains nominal bias and credible interval coverage across a wide range of tail shapes, whereas the SL counterpart performs similarly only when its Laplace tails are correct and loses accuracy when a heavy, misspecified tail dominates the uncensored data.

    A complete step-by-step implementation guide, containing annotated \texttt{JAGS} and \texttt{R} code, as well as the full script that reproduces the ACTG~315 HIV viral load analysis, is available in the online repository.

    The remainder of the article is organized as follows. \autoref{sec:DATASET} introduces the motivating data. \autoref{sec:QUANTDIST} reviews the SL and SEP distributions, discusses their flexibility, and shows how each handles censoring. \autoref{sec:GENERAL_MODEL} presents the hierarchical model and prior choices. \autoref{sec:COMPARE_ADEQUACY} outlines model-checking procedures, \autoref{sec:APPLICATION} reports the data analysis, and \autoref{sec:SIMULATION} summarizes simulation results. We conclude in \autoref{sec:DISCUSS}.

    \section{Illustrative case study}\label{sec:DATASET}

    The motivating data come from the AIDS Clinical Trials Group (ACTG) 315 study, which followed 46 participants on antiretroviral therapy for 196 days. Plasma viral load was measured repeatedly, and 11\% of the 514 observations were left-censored at the assay's lower limit of detection (LLD) of 100~copies/mL ($\log_{10}=2$). The dataset is publicly available in the \texttt{R} package \texttt{ushr} \citep{MORRIS2020}.

\begin{figure}
      \centering
      \includegraphics[width=0.7\textwidth]{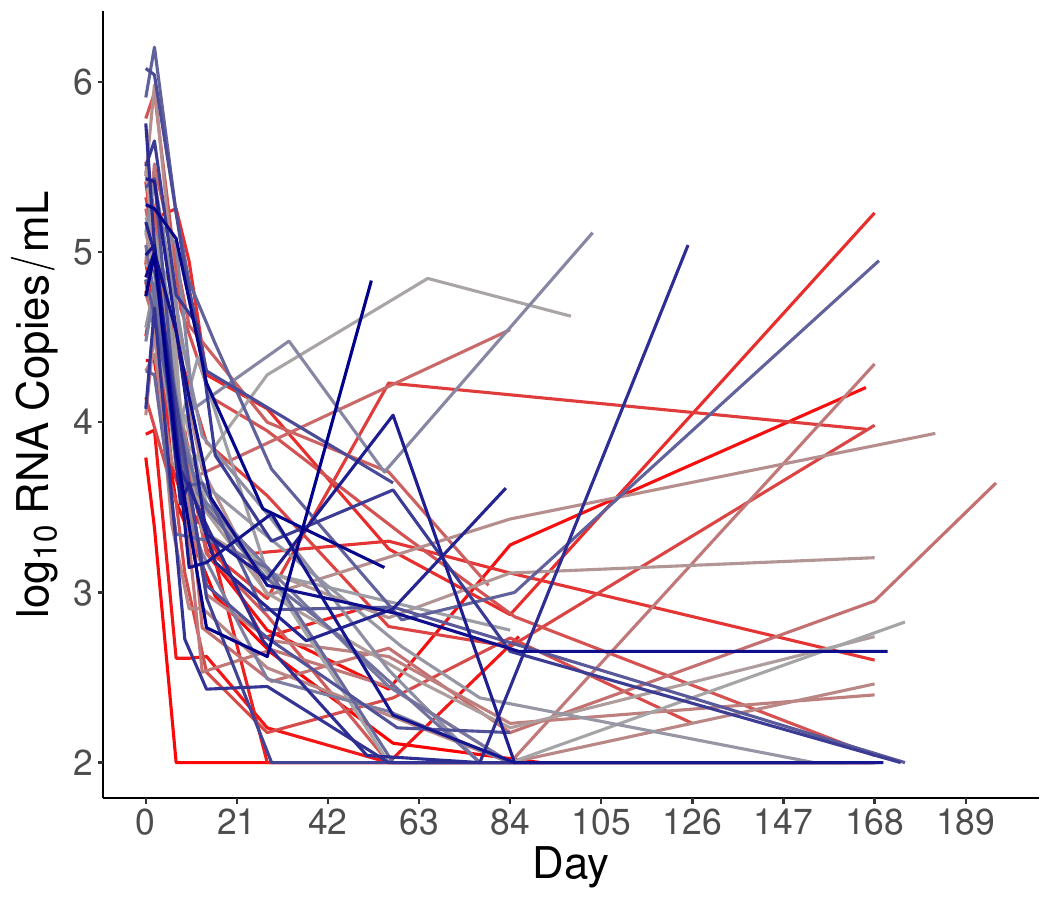}
      \caption{Longitudinal $\log_{10}$ viral load trajectories for 46 ACTG~315 participants. The censoring limit is 100~copies/mL ($\log_{10}=2$).}
      \label{fig:ACTG315_RNA}
    \end{figure}

    Under potent therapy, HIV RNA typically shows a biphasic (biexponential) decline: an initial rapid phase driven by the clearance of short-lived infected cells, followed by a slower phase reflecting longer-lived reservoirs \citep{perelson1996hiv}. Variation in these phases is clinically relevant because upper-tail trajectories can signal sub-optimal suppression or emerging resistance. \autoref{fig:ACTG315_RNA} plots the $\log_{10}$ viral load profiles for all participants; censored observations are displayed at the LLD. While most trajectories remain clustered near the center, several patients' curves rise sharply in the later study period, resulting in a distinctly right-skewed distribution of viral loads.

    Mean-based mixed models can hide clinically important heterogeneity because they describe only the distribution's center. Fitting several prespecified quantiles reveals how covariates affect patients in the upper and lower tails, for example, individuals with unusually slow decay or exceptionally rapid suppression. However, an SL error term enforces left skewness for high quantile levels ($p_{0}>0.5$); therefore, it is structurally mismatched to the right-skewed upper tail visible in \autoref{fig:ACTG315_RNA} and risks systematic misfit. To accommodate the heavy right tail and left-censoring, we adopt the SEP likelihood and benchmark it against the SL alternative. (The same likelihood form extends to right- or interval-censoring, though those cases do not occur here.)

    Earlier joint models for viral load and CD4 counts included explicit measurement error components and focused on mean regression with lighter-tailed likelihoods (e.g., see Wu \citep{wu2002joint}). In line with Yu and Yu \citep{yu2023flexible} and Burger et al. \citep{burger2025robust}, our analysis treats viral load as the sole response and uses CD4 count only as a time-varying covariate. We maintain a two-level hierarchy in which patient-specific random effects drive a biphasic (sum-of-exponentials) decay curve characterized by four population-level parameters: two initial loads and two clearance rates. A specific biphasic decay form is given in \autoref{sec:NLME_MODELS}
    
    Because censoring is one-sided here, the CDF contribution enters the likelihood in a particularly simple way, but the framework could accommodate more complex censoring patterns with the same analytic CDF.

    \section{Quantile-parameterized distributions} \label{sec:QUANTDIST}

    We now formalize the two error distributions that underlie our QMM: the SL and the SEP. The SL family is popular in parametric quantile regression because its location parameter coincides with the target quantile. However, its skewness and tail thickness are fixed once that quantile is chosen. The SEP family relaxes this constraint by introducing two tail parameters, which allow the left and right tails to vary independently. Below, we review their probability density functions (PDFs) and show how each handles censoring through its CDF.

    \subsection{Skew Laplace distribution}
    
    Let $\mu$, $\sigma$, and $p_0$ specify the SL distribution, where $\mu$ is set to correspond to the $p_0$\textsuperscript{th} quantile, $\sigma > 0$ is the scale parameter, and $p_0 \in \left(0,1\right)$ governs the skewness \citep{GERACI2007}.
    
    The SL PDF is given by:
    \begin{equation}
        f_{\text{SL}} \left(y \left| \mu, \sigma, p_0 \right.\right) = 
        \begin{cases}
        \displaystyle \frac{2 p_0 \left(1 - p_0\right)}{\sigma} \exp\left(\frac{2\left(y - \mu\right)\left(1 - p_0\right)}{\sigma}\right), & \text{if } y \leq \mu, \\
        \displaystyle \frac{2 p_0 \left(1 - p_0\right)}{\sigma} \exp\left(-\frac{2 p_0 \left(y - \mu\right)}{\sigma}\right), & \text{if } y > \mu.
        \end{cases}
    \end{equation}
    The corresponding CDF is:
    \begin{equation}
        F_{\text{SL}} \left(y \left| \mu, \sigma, p_0 \right.\right) =
        \begin{cases}
        \displaystyle p_0 \exp\left(\frac{2\left(y - \mu\right)\left(1 - p_0\right)}{\sigma}\right), & \text{if } y \leq \mu, \\
        \displaystyle 1 - \left(1 - p_0\right) \exp\left(-\frac{2 p_0 \left(y - \mu\right)}{\sigma}\right), & \text{if } y > \mu.
        \end{cases}
    \end{equation}
    The SL distribution has moderately heavy tails, and it does not include a separate parameter for adjusting tail behavior.
    
    \subsection{Skew exponential power distribution}
    
    To achieve greater flexibility in both skewness and tail thickness while retaining an analytic CDF, we adopt the SEP distribution. Let $\mu$ denote the $p_0$\textsuperscript{th} quantile ($0<p_{0}<1$), $\sigma>0$ the scale, and $\kappa_{1},\kappa_{2}>0$ the left- and right-tail shape parameters. Following Zhu and Zinde-Walsh \citep{zhu2009properties}, the PDF is
    \begin{equation}
        f_{\text{SEP}}\left(y \left| \mu, \sigma, \kappa_1, \kappa_2, p_0 \right.\right) =
        \begin{cases}
            \frac{1}{\sigma} \exp\left(-\frac{1}{\kappa_1}\left(\frac{\mu - y}{2p_0\sigma K_1}\right)^{\kappa_1}\right), & y \leq \mu, \\
            \frac{1}{\sigma} \exp\left(-\frac{1}{\kappa_2}\left(\frac{y - \mu}{2\left(1 - p_0\right)\sigma K_2}\right)^{\kappa_2}\right), & y > \mu,
        \end{cases}
    \end{equation}
    where the normalizing constants are
    \begin{equation} \label{eq:K1_K2}
        K_j = \frac{\kappa_j^{-1/\kappa_j}}{2 \Gamma\left(1 + \frac{1}{\kappa_j}\right)}, \quad j=1,2.
    \end{equation}
    The distribution satisfies:
    \begin{equation}
        \int_{-\infty}^{\mu}f_{\text{SEP}}\left(y \left| \mu, \sigma, \kappa_1, \kappa_2, p_0 \right.\right)\mathrm{d}y = p_0,
    \end{equation}
    thus defining $\mu$ as the $p_0$\textsuperscript{th} quantile.

    For $\kappa_{1}<2$, the left tail of the SEP distribution is heavier than a normal tail, and for $\kappa_{2}<2$, its right tail is heavier; when either shape parameter exceeds 2, the corresponding tail becomes lighter than the normal tail (with $\kappa_{1}=\kappa_{2}=2$ reproducing normal tails on both sides).
    
    A corrected CDF derivation appears in the Appendix; it fixes a scaling error in the incomplete gamma term of the expression reported by  Zhu and Zinde-Walsh \citep{zhu2009properties}. The resulting CDF is
    \begin{equation}
        F_{\text{SEP}}\left(y \left| \mu, \sigma, \kappa_1, \kappa_2, p_0 \right.\right) =
        \begin{cases}
            p_0 \left(1 - G\left(\frac{1}{\kappa_1}\left(\frac{\mu - y}{2 p_0 \sigma K_1}\right)^{\kappa_1}, \frac{1}{\kappa_1}\right)\right), & y \leq \mu, \\
            p_0 + \left(1 - p_0\right) G\left(\frac{1}{\kappa_2}\left(\frac{y - \mu}{2\left(1 - p_0\right) \sigma K_2}\right)^{\kappa_2}, \frac{1}{\kappa_2}\right), & y > \mu,
        \end{cases}
    \end{equation}
    where $G\left(a, b\right)$ is the regularized incomplete gamma function:
    \begin{equation}
        G\left(a, b\right) = \frac{1}{\Gamma\left(b\right)} \int_{0}^{a} t^{b-1} e^{-t} \mathrm{d}t,
    \end{equation}
    with $\Gamma\left(b\right)$ as the standard gamma function:
    \begin{equation}
        \Gamma\left(b\right) = \int_{0}^{\infty} t^{b-1} e^{-t}\mathrm{d}t.
    \end{equation}
    The SEP distribution reduces to a normal distribution with mean $\mu$ and standard deviation $\sigma/\sqrt{2\pi}$ when $p_0 = 0.5$ and $\kappa_1 = \kappa_2 = 2$. When $p_0 = 0.5$ and $\kappa_1 = \kappa_2 = 1$, it becomes the symmetric Laplace distribution. As $\left(\kappa_1, \kappa_2\right) \to \infty$, the SEP distribution approaches a uniform distribution.

    \begin{figure}[h!]
        \centering
        \begin{subfigure}{\textwidth}
            \centering
            \includegraphics[width=0.45\textwidth]{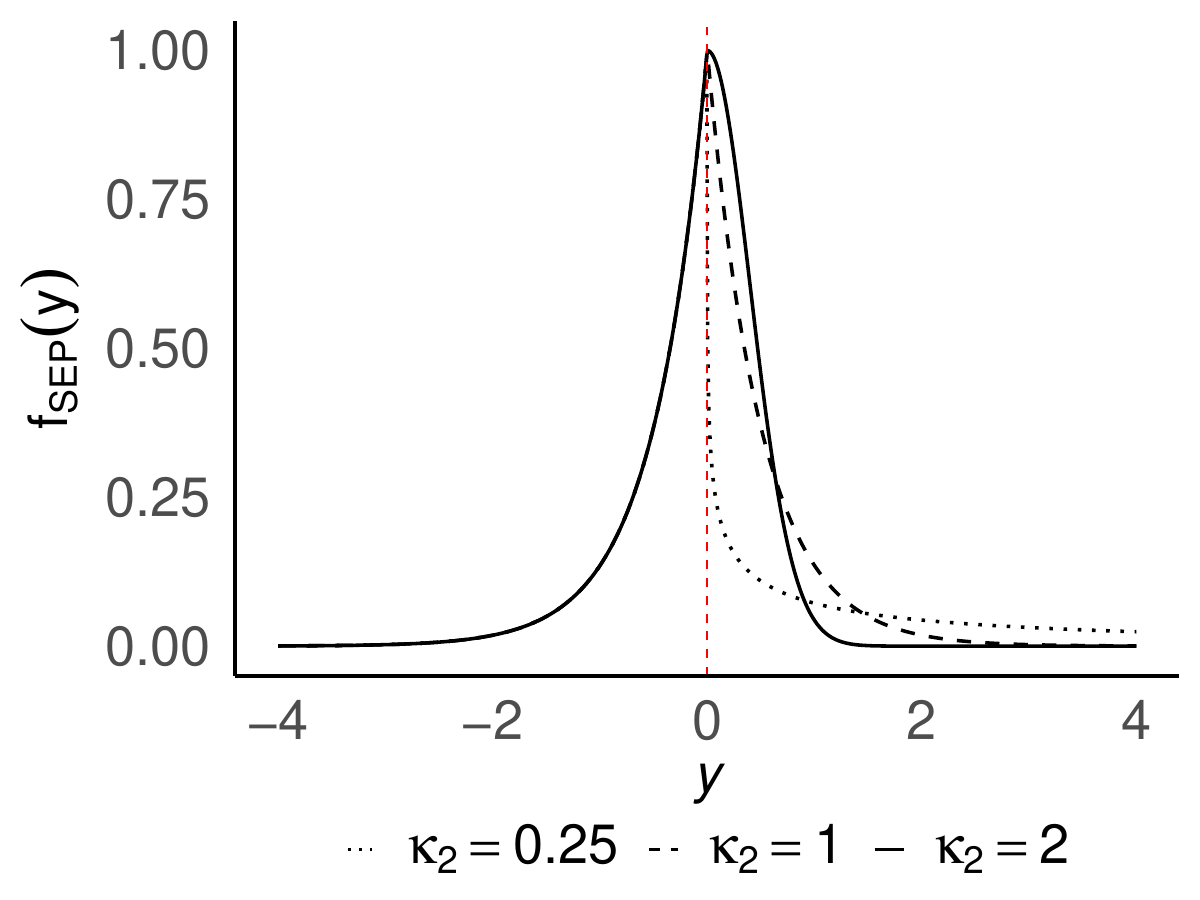}\hspace{-0.4cm}
            \includegraphics[width=0.45\textwidth]{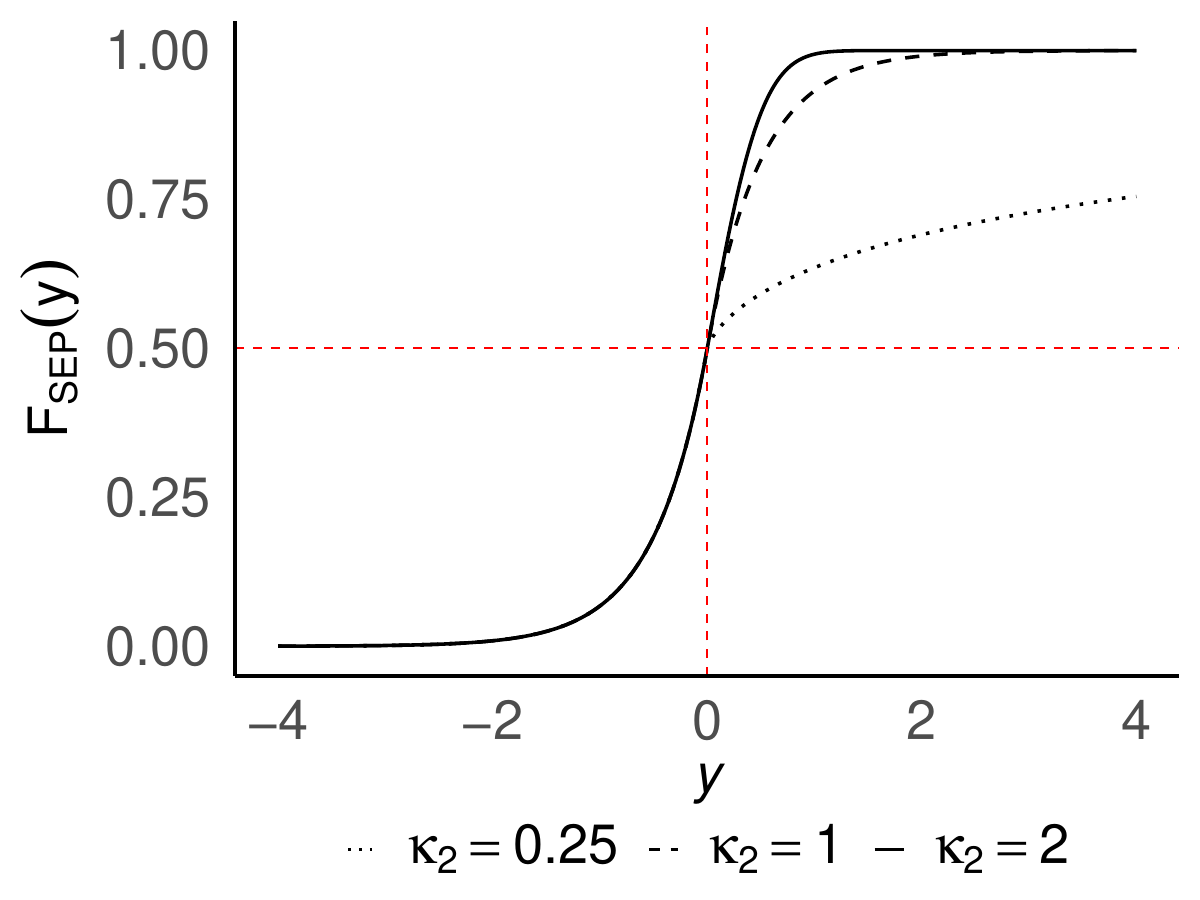}
            \caption{$p_0 = 0.5$}
        \end{subfigure}
        \vspace{0.5cm}
        \begin{subfigure}{\textwidth}
            \centering
            \includegraphics[width=0.45\textwidth]{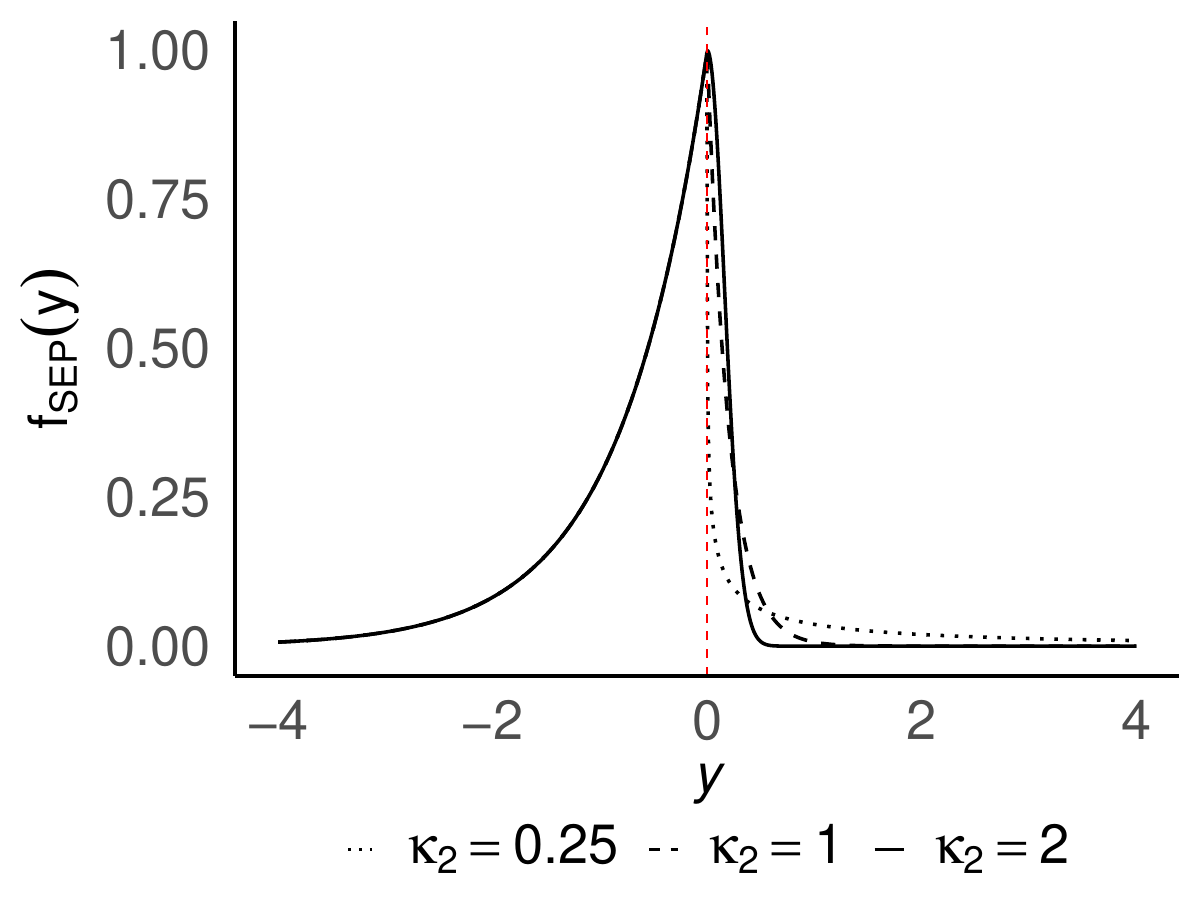}\hspace{-0.4cm}
            \includegraphics[width=0.45\textwidth]{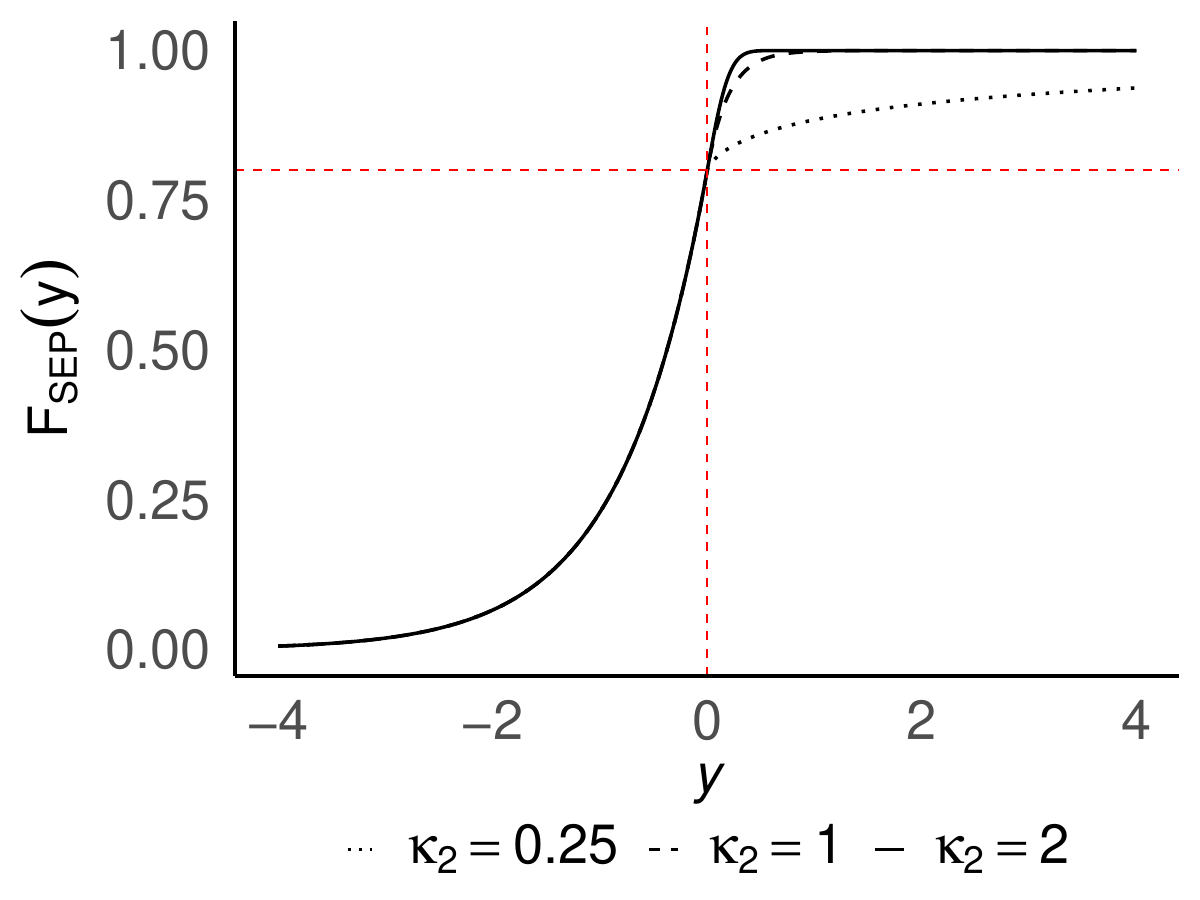}
            \caption{$p_0 = 0.8$}
        \end{subfigure}
        \caption{PDF (left) and CDF (right) of the SEP distribution for two quantile levels ($p_{0}=0.5$ and $0.8$; rows). Parameters are fixed at $\mu=0$, $\sigma=1$, $\kappa_{1}=1$, while the right-tail shape varies $\kappa_{2}\in\left\{0.25\;\left(\text{dotted}\right),\,1\;\left(\text{dashed}\right),\,2\;\left(\text{solid}\right)\right\}$. Vertical and horizontal red dashed lines highlight the $p_0$\textsuperscript{th} quantile $\left(\mu,p_{0}\right)$. The black dashed curve is the SL special case $\kappa_{1}=\kappa_{2}=1$. Smaller $\kappa_{2}$ values thicken the right tail; larger values thin it.}
        \label{fig:SEP_PDF_CDF}
    \end{figure}
    
    In \autoref{fig:SEP_PDF_CDF}, the PDF and CDF of the SEP distribution are shown at two quantile levels ($p_0 = 0.5$ and $0.8$) for different values of $\kappa_2$, with $\mu = 0$, $\sigma = 1$, and $\kappa_1 = 1$ held fixed. The parameter $p_0$ determines the location of the $p_0^\text{th}$ quantile, anchoring it at $\mu$. When $p_0 > 0.5$, the distribution places more mass to the left of $\mu$, resulting in left-skewness; when $p_0 < 0.5$, more mass is shifted to the right, producing right-skewness. Varying $p_0$ allows direct control over the position and skew of the distribution. In addition, the shape parameters $\kappa_1$ and $\kappa_2$ control tail behavior on each side of $\mu$: smaller values lead to heavier, more slowly decaying tails, while larger values produce lighter, thinner tails.

    \section{General modeling framework} \label{sec:GENERAL_MODEL}

    Suppose we have $N$ subjects, indexed by $i = 1,\dots,N$. For each subject, we collect $J_i$ repeated measurements. Let $y_{ij}$ denote the outcome of interest (e.g., a biomarker measurement) for subject $i$ at time point $j$. To accommodate (left) censoring, define an indicator $\delta_{ij} \in \left\{0,1\right\}$, where $\delta_{ij} = 1$ indicates that $y_{ij}$ is fully observed, and $\delta_{ij} = 0$ indicates censoring below a detection threshold.

    \subsection{Quantile model}

    We aim to model the $p_0$\textsuperscript{th} quantile of $y_{ij}$, denoted $\mu_{ij}\left(p_0,\mathbf{v}_i,\bm{\beta}\right)$. Here, $p_0 \in \left(0,1\right)$ is a user-specified level rather than an estimable parameter.

    To capture inter-subject variability, we introduce a subject-level random effect vector $\mathbf{v}_i$, assumed to follow a multivariate normal distribution:
    \begin{equation}
      \mathbf{v}_i \sim \mathcal{N} \left(\mathbf{0}, \bm{\Sigma}_v\right),
    \end{equation}
    where $\bm{\Sigma}_v$ is an unknown covariance matrix. Let $\mathbf{x}_{ij}$ denote the covariate vector for subject $i$ at time point $j$ (the observation time $t_{ij}$ is included as one of the covariates). Then the $p_0$\textsuperscript{th} quantile is modeled as
    \begin{equation}
      \mu_{ij}\left(p_0,\mathbf{v}_i,\bm{\beta}\right) = g\left(\mathbf{x}_{ij}, \mathbf{v}_i, \bm{\beta}\right),
    \end{equation}
    where $g\left(\cdot\right)$ is a user-defined linear or nonlinear link, and $\bm\beta$ collects the fixed effects. We choose $g\left(\cdot\right)$ so that $\mu_{ij}\left(p_0,\mathbf{v}_i,\bm{\beta}\right)$ is, by construction, the conditional $p_0$\textsuperscript{th} quantile under either the SL or SEP error model.

    \subsection{Handling censoring and likelihood}

    Given $\mu_{ij}\left(p_0,\mathbf{v}_i,\bm{\beta}\right)$, we assume $y_{ij}$ follows either the SL or SEP distribution, parameterized by $\left(\mu_{ij}\left(p_0,\mathbf{v}_i,\bm{\beta}\right),\sigma\right)$ and, in the SEP case, additional tail parameters $\left(\kappa_1,\kappa_2\right)$. 
    Let $f_{\mathrm{dist}}\left(\cdot\right)$ and $F_{\mathrm{dist}}\left(\cdot\right)$ denote the chosen PDF and CDF, respectively.
    
    Conditional on the random effects $\mathbf{v}_i$, the likelihood contribution of $\left(y_{ij},\delta_{ij}\right)$ is
    \begin{equation} \label{eq:like_ij}
      \ell_{ij}\left(\bm{\Theta}\right)
      =
      \left[f_{\mathrm{dist}}\left(y_{ij}\left|\mu_{ij}\left(p_0,\mathbf{v}_i,\bm{\beta}\right),\dots\right.\right)\right]^{\delta_{ij}}
      \left[F_{\mathrm{dist}}\left(y_{ij}\left|\mu_{ij}\left(p_0,\mathbf{v}_i,\bm{\beta}\right),\dots\right.\right)\right]^{1-\delta_{ij}},
    \end{equation}
    so fully observed outcomes use the PDF, whereas left-censored outcomes enter through the CDF. Aggregating over all subjects and times,
    \begin{equation} \label{eq:likelihood}
      \mathcal{L}\left(\bm{\Theta}\right)
      =
      \prod_{i=1}^{N}\prod_{j=1}^{J_i}\ell_{ij}\left(\bm{\Theta}\right),
    \end{equation}
    where $\bm{\Theta}$ collects all unknowns, namely $\bm{\beta}$, $\left\{\mathbf{v}_i\right\}$, $\bm{\Sigma}_v$, and any SL/SEP-specific parameters (e.g., $\sigma,\kappa_1,\kappa_2$).

    The SL kernel also has a normal-exponential mixture representation, which allows censored observations to be handled via a Tobit-style data augmentation step \citep{kozumi2011gibbs}. We nonetheless retain the analytic CDF approach for both models to maintain parallel treatment, since no analogous hierarchical (variance-mixture) representation is available for the more flexible SEP kernel.
    
    Although we focus on left censoring here, right or interval censoring can be accommodated by modifying \eqref{eq:likelihood} to involve $F_{\mathrm{dist}}$ or its complement as appropriate.

    \subsection{Prior specification}

    We assign weakly informative priors to all parameters. The fixed effects vector $\bm{\beta}$ receives a diffuse multivariate normal prior
    \begin{equation}\label{eq:beta_prior}
      \bm{\beta}
      \sim
      \mathcal{N}\left(\mathbf{0},\sigma_\beta^2\mathbf{I}\right),
    \end{equation}
    where $\sigma_\beta^2$ is a large variance constant. For the random effects covariance $\bm{\Sigma}_v$, we use a Cholesky factor $\mathbf{L}_v$ such that $\bm{\Sigma}^{-1}_v=\mathbf{L}_v\mathbf{L}_v^\top$ (factorizing the precision matrix). Each diagonal entry of $\mathbf{L}_v$ has a half-$t$ (or similar) prior, while off-diagonals may have independent diffuse, normal priors; specific hyperparameters can be chosen to suit the application.

    Both the SL and SEP include a global scale parameter $\sigma$, to which we assign a half-$t$ prior. In the SEP formulation, we add tail parameters $\kappa_1$ and $\kappa_2$, each given a similarly weak prior truncated to $\left(0,\infty\right)$. Subject-matter knowledge might justify constraints such as $\kappa_i \le 2$ to avoid extremely light tails, but in general, we allow the data to guide inference on tail behavior. For instance, Bernardi et al. \citep{bernardi2018bayesian} assign a $\mathrm{Uniform}\left(0,2\right)$ prior to the shared tail parameter $\kappa$ ($\kappa_{1} = \kappa_{2}$). This ensures the model's tails are at least as heavy as those of a normal distribution, providing robustness against outliers.

    Combining the likelihood in \eqref{eq:likelihood} with these generic priors yields the joint posterior distribution of all parameters, which can be sampled via Markov chain Monte Carlo (MCMC). In the following sections, we illustrate this framework for the HIV viral load data and compare the SL and SEP models under left censoring.

    \section{Model comparison and adequacy} \label{sec:COMPARE_ADEQUACY}

    Having described the hierarchical model structure and prior specification, we now outline our approach to model comparison via marginal likelihood, followed by diagnostic checks based on posterior predictive residuals.

    \subsection{Marginal likelihood} \label{sec:MARGINAL_LIKE}
    
    Our approach is hierarchical with subject-level random effects. For model comparison, we use the marginal likelihood, which we obtain by integrating out the random effects instead of conditioning on them \citep{gelman2013bayesian}. Many Bayesian mixed-effects analyses assess model fit by conditioning on subject-specific effects. A common tool is the ``conditional'' deviance information criterion of Spiegelhalter et al. \citep{SPIEGELHALTER2002}, appreciated for its straightforward calculation from typical MCMC output. However, treating the latent random effects as fixed parameters ignores their prior uncertainty and ultimately understates model complexity \citep{QUINTERO2018}. The marginal likelihood penalizes that complexity automatically \citep{deSantis1998bayes}. Because the resulting high-dimensional integral is analytically intractable, we compute it with bridge sampling algorithms, which give accurate numerical estimates \citep{gronau2020computing}.

    After fitting each model in \texttt{JAGS}, we obtain posterior draws $\{\bm{\Theta}_{s}\}_{s=1}^{S}$, where each $\bm{\Theta}_{s}$ contains all unknown quantities, including the subject-level random effects $\mathbf{v}_{i}$. For every draw, we evaluate the unnormalized posterior density
    \begin{equation}
      p\left(\bm{\Theta}_{s}\left|\text{data}\right.\right) \propto 
      p\left(\text{data}\left|\bm{\Theta}_{s}\right.\right)
      p\left(\bm{\Theta}_{s}\right) \coloneqq \tilde{p}\left(\bm{\Theta}_{s}\left|\text{data}\right.\right).
    \end{equation}
    The \texttt{bridgesampling} package \citep{JSSv092i10} combines these unnormalized values with an automatically generated proposal distribution in a bridge sampling scheme to solve for the log marginal likelihood $\log p\left(\text{data}\right)$. This quantity serves as a coherent basis for model comparison, as it incorporates both the hierarchical structure and prior information.

    \subsection{Residual diagnostics}

    We assess model adequacy for uncensored observations using simulation-based residual diagnostics. At each MCMC iteration, we proceed as follows. First, we take the current draws of the global parameters, including the fixed effects vector $\bm{\beta}$, the covariance matrix $\bm{\Sigma}_v$ of the subject-level random effects, and the residual scale and shape parameters of the observation model. Using these draws, we generate a new set of random effects
    \begin{equation}
        \tilde{\mathbf{v}}_i \sim \mathcal{N}\left(\mathbf{0}, \bm{\Sigma}_v\right)
    \end{equation}
    for all individuals $i$, rather than reusing the $\mathbf{v}_i$ estimated from the observed data.

    Second, for each observed covariate vector $\mathbf{x}_{ij}$, we evaluate the model-implied conditional quantile
    \begin{equation}
        \mu_{ij}\left(p_0,\tilde{\mathbf{v}}_i,\bm{\beta}\right)
        = g\!\left(\mathbf{x}_{ij}, \tilde{\mathbf{v}}_i, \bm{\beta}\right).
    \end{equation}
    Third, we draw replicated outcomes $\tilde{y}_{ij}$ from the assumed observation model at the specified quantile level (SL or SEP), using $\mu_{ij}\left(p_0,\tilde{\mathbf{v}}_i,\bm{\beta}\right)$ as the location together with the current residual scale and shape parameters. Across posterior draws, this yields a large collection of simulated datasets that represent what the fitted model predicts after integrating over both the random effects distribution and the observation-level noise.

    We then restrict attention to rows that were uncensored in the original data and pass the resulting matrix of simulated $\tilde{y}_{ij}$ values, together with the corresponding observed $y_{ij}$ values, to the \texttt{DHARMa} package \citep{HARTIG2021A,HARTIG2021B}. \texttt{DHARMa} converts these simulations into scaled residuals by comparing each observation to its own marginal posterior predictive distribution. Under a well-calibrated model, these scaled residuals should be approximately $\text{Uniform}\left(0,1\right)$.

    We assess the uniformity of the scaled residuals using QQ plots that compare their empirical quantiles with the expected uniform quantiles. Points that follow the diagonal indicate an adequate joint specification of the fixed effects, the random effects distribution, and the chosen error distribution. Systematic departures, for example, S-shaped curves, suggest a lack of fit in one or more of these components.

    \section{Application to HIV dataset} \label{sec:APPLICATION}

    We fit a biphasic, nonlinear QMM with SEP- and SL-distributed errors, as detailed next.

    \subsection{Nonlinear mixed-effects model} \label{sec:NLME_MODELS}

    We model each patient's viral load trajectory using a biphasic (two-phase) decay function, allowing left-censoring below 100~copies/mL and permitting flexible quantile-based inference. In addition to time, we include the observed CD4 count, $\mathrm{cd4}_{ij}$, as a time-varying covariate that modifies the slower decay phase.

    Let $y_{ij}$ be the viral load measurement on the $\log_{10}$ scale for subject $i$ at time $t_{ij}$, where $i=1,\dots,N$ and $j=1,\dots,J_i$. Define $\delta_{ij}\in\left\{0,1\right\}$ as an indicator that $y_{ij}$ is fully observed $\left(\delta_{ij}=1\right)$ or left censored $\left(\delta_{ij}=0\right)$.

    To capture subject-level variation, each patient $i$ has a four-dimensional random effects vector
    \begin{equation}\label{eq:v_i}
      \mathbf{v}_i
      =
      \left(v_{i,1},v_{i,2},v_{i,3},v_{i,4}\right)^{\top}
      \sim
      \mathcal{N}\left(\bm{0},\bm{\Sigma}_v\right),
    \end{equation}
    where $\bm{\Sigma}_v$ is a $4\times4$ covariance matrix.

    We map the random effects $\left(v_{i,1},v_{i,2},v_{i,3},v_{i,4}\right)$ onto a biexponential decay model with a fast and a slow phase. For patient $i$,
    \begin{equation}\label{eq:bi_exp}
        \begin{aligned}
            P_1^{\left(i\right)} &= \exp\left(\beta_1 + v_{i,1}\right), & \lambda_1^{\left(i\right)} &= \beta_2 + v_{i,2}, \\
            P_2^{\left(i\right)} &= \exp\left(\beta_3 + v_{i,3}\right), & \lambda_2^{\left(ij\right)} &= \beta_4 + v_{i,4}+\gamma\mathrm{cd4}_{ij},
        \end{aligned}
    \end{equation}
    so that $P_1^{\left(i\right)}$ is the initial viral load of the short-lived, rapidly clearing compartment with clearance rate $\lambda_1^{\left(i\right)}$, whereas $P_2^{\left(i\right)}$ is the initial load of the long-lived reservoir whose slower clearance rate $\lambda_2^{\left(ij\right)}$ is adjusted at each time point $j$ by the corresponding CD4 count through the coefficient $\gamma$. The fixed effects vector $\bm{\beta}=\left(\beta_1,\dots,\beta_4\right)^{\top}$ sets the population-level coefficients for the $p_{0}$\textsuperscript{th} quantile model, and the terms $v_{i,k}$ capture patient-specific deviations from those coefficients.

    Define
    \begin{equation}\label{eq:mu_ij}
      \mu_{ij}\left(p_0,\mathbf{v}_i,\bm{\beta}\right)
      =
      \log_{10}\left(
        P_1^{\left(i\right)}\exp\left(-\lambda_1^{\left(i\right)}t_{ij}\right)+
        P_2^{\left(i\right)}\exp\left(-\lambda_2^{\left(ij\right)}t_{ij}\right)
     \right),
    \end{equation}
    so $\mu_{ij}\left(p_0,\mathbf{v}_i,\bm{\beta}\right)$ is the $p_0$\textsuperscript{th} quantile under either SL- or SEP-distributed errors. We refer to the two specifications as SKL-QMM and SEP-QMM, respectively.

    Given $\mathbf{v}_i$, each data pair $\left(y_{ij},\delta_{ij}\right)$ contributes the usual censored-data term: the density if $\delta_{ij}=1$ and the CDF if $\delta_{ij}=0$; see \eqref{eq:like_ij} for the exact expression. Let $\bm{\Theta}$ denote the full set of unknown parameters.
  
    We assign independent $\mathcal{N}\left(0,1000\right)$ priors to $\beta_{1},\dots,\beta_{4}$ and to $\gamma$. For the random effects covariance, we write $\bm{\Sigma}_v=\left(\mathbf{L}_v\mathbf{L}_v^{\top}\right)^{-1}$, where $\mathbf{L}_v$ is a $4\times4$ Cholesky factor. Its diagonal elements have independent half-$t$ priors, $L_{v,ii}\sim t_3^{+}\left(0,\sqrt{2}\right)$ \citep{GELMAN2006A} and its off-diagonal elements follow $L_{v,ij}\sim\mathcal{N}\left(0,1000\right)$ for $i>j$, thereby giving a weakly informative prior on the correlations among $\left(P_1^{\left(i\right)},\lambda_1^{\left(i\right)},P_2^{\left(i\right)},\lambda_2^{\left(ij\right)}\right)$. Here, $t_\nu^{+}\left(0,s\right)$ denotes a half-$t$ distribution with $\nu$ degrees of freedom, location $0$, and scale $s$.

    Both the SL and SEP models include a global scale parameter with $\sigma\sim t_3^{+}\left(0,\sqrt{2}\right)$, and in the SEP model each tail parameter has the same half-$t$ prior $\kappa_1,\kappa_2\sim t_3^{+}\left(0,\sqrt{2}\right)$.

    Combining these priors with the likelihood $\prod_{i=1}^{N}\prod_{j=1}^{J_i}\ell_{ij}\left(\bm{\Theta}\right)$ gives the joint posterior distribution of $\bm{\Theta}$. We sample from this posterior via MCMC, implemented in \texttt{JAGS}, and base all parameter estimates on the resulting draws.
    
    \subsection{Model implementation}

    We fit SKL-QMM and SEP-QMM to the ACTG~315 data using the hierarchical framework in \autoref{sec:NLME_MODELS}, estimating both models at five quantiles $p_0 \in \left\{0.10, 0.25, 0.50, 0.75, 0.90\right\}$ to cover the lower, median, and upper parts of the viral load distribution. Posterior sampling was carried out in \texttt{JAGS} \citep{PLUMMER2003}, with convergence monitored via the potential scale reduction factor ($\hat{R}$). We estimated marginal likelihoods via bridge sampling (\texttt{bridgesampling}; \citep{gronau2020computing}) and assessed fit with posterior predictive residual checks in \texttt{DHARMa} \citep{HARTIG2021A}, as outlined in \autoref{sec:COMPARE_ADEQUACY}. We report posterior medians as point estimates and use 95\% Bayesian credible intervals (BCIs) to quantify uncertainty. All code is provided on \href{https://github.com/DABURGER1/Censored-SEP-Mixed-Model}{GitHub}, and computations were run on a system with an 11\textsuperscript{th}-generation Intel\textsuperscript{\textregistered} Core\textsuperscript{TM} i5-1145G7 processor (2.60~GHz) with 16~GB of RAM.

    \subsection{Model comparison}

    \autoref{tab:MODEL_COMPARISON} shows the log-marginal likelihoods for SKL-QMM and SEP-QMM across the five quantiles ($p_{0}=0.10\!-\!0.90$); larger values indicate a better fit. SEP-QMM outperforms SKL-QMM at every quantile, with log-likelihood gaps (SEP minus SL) ranging from $+3.1$ at $p_{0}=0.10$ to $+20.5$ at $p_{0}=0.75$. On the Kass-Raftery scale, differences above $2$ are considered positive evidence, and those above $10$ are deemed decisive \citep{Kass1995}. So, these results strongly favor the SEP, especially for the higher-quantile models. The advantage arises because the SEP can thicken or thin each tail independently of $p_{0}$, allowing it to match the pronounced right skew of high viral loads, whereas the SL is locked into a left skew at those levels.

    \setlength{\tabcolsep}{6pt}
    \begin{longtable}{lS[table-format=4.2]S[table-format=4.2]}
        \caption{Log-marginal likelihoods for SKL-QMM and SEP-QMM; larger values indicate a better fit.}\label{tab:MODEL_COMPARISON}\\
        \toprule
        & \multicolumn{2}{c}{\textbf{Model}}\\
        \cline{2-3}
        $\bm{p_0}$ & \textbf{SKL-QMM} & \textbf{SEP-QMM}\\
        \midrule
        \endfirsthead
    
        \multicolumn{3}{l}{\textit{Continued from previous page}}\\
        \toprule
        \textbf{Quantile} & \textbf{SKL-QMM} & \textbf{SEP-QMM}\\
        \midrule
        \endhead
    
        \toprule
        \multicolumn{3}{r}{\textit{Continued on next page}}\\
        \endfoot
    
        \bottomrule
        \endlastfoot
    
        0.10 & -330.71 & -327.56\\*
        0.25 & -349.29 & -341.64\\*
        0.50 & -339.29 & -331.17\\*
        0.75 & -358.15 & -337.70\\*
        0.90 & -359.96 & -345.16\\
    \end{longtable}

    \subsection{Regression fits}

    With SEP-QMM favored by the marginal likelihood, we next compare its population-level quantile trajectories with those from SKL-QMM to see how the choice of error distribution shapes the fitted curves.

\begin{figure}
        \centering
        \begin{subfigure}{0.45\textwidth}
            \centering
            \includegraphics[width=\linewidth]{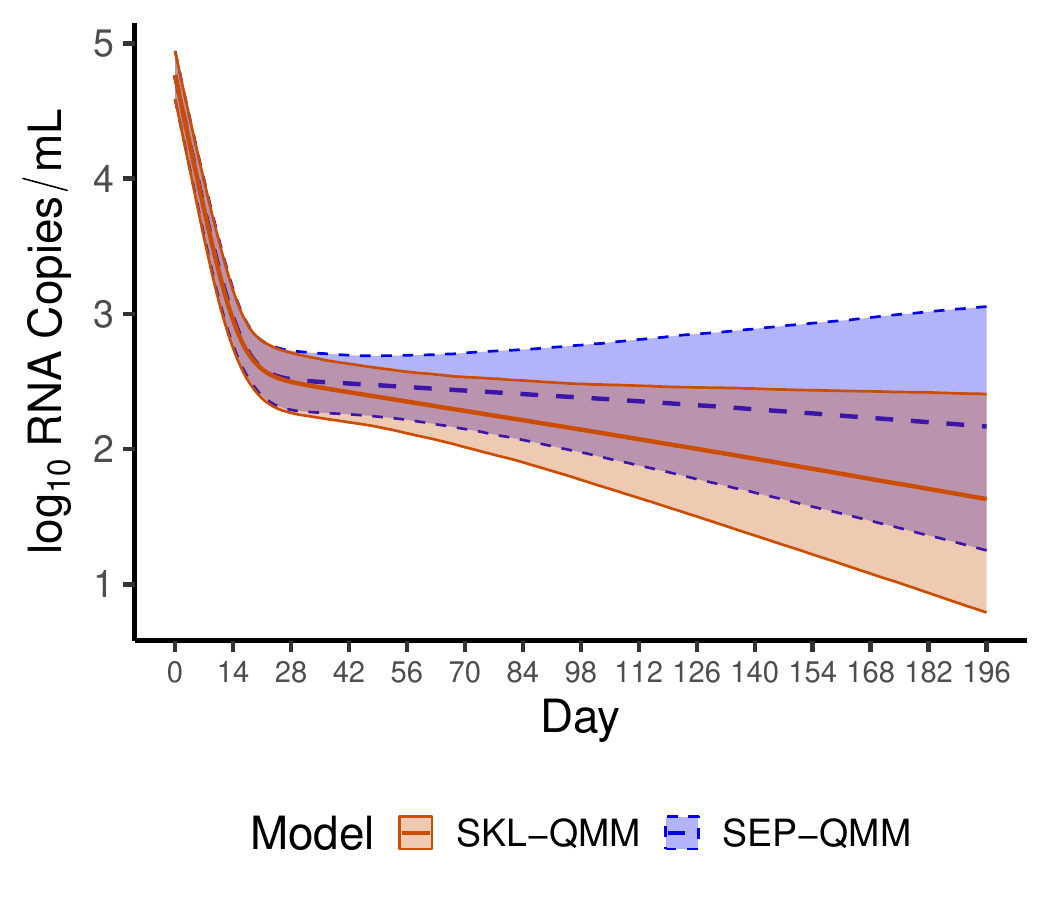}
            \caption{$p_0 = 0.10$}
        \end{subfigure}
        \begin{subfigure}{0.45\textwidth}
            \centering
            \includegraphics[width=\linewidth]{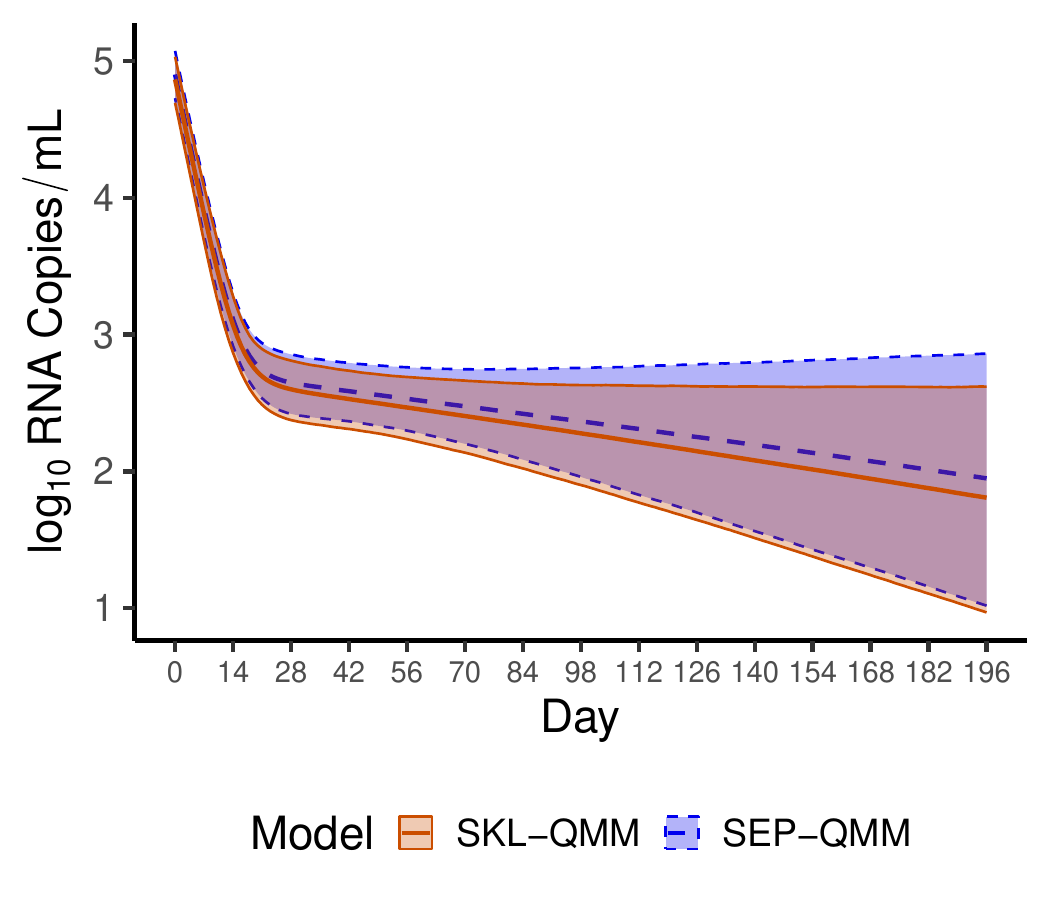}
            \caption{$p_0 = 0.25$}
        \end{subfigure}
        \\
        \begin{subfigure}{0.45\textwidth}
            \centering
            \includegraphics[width=\linewidth]{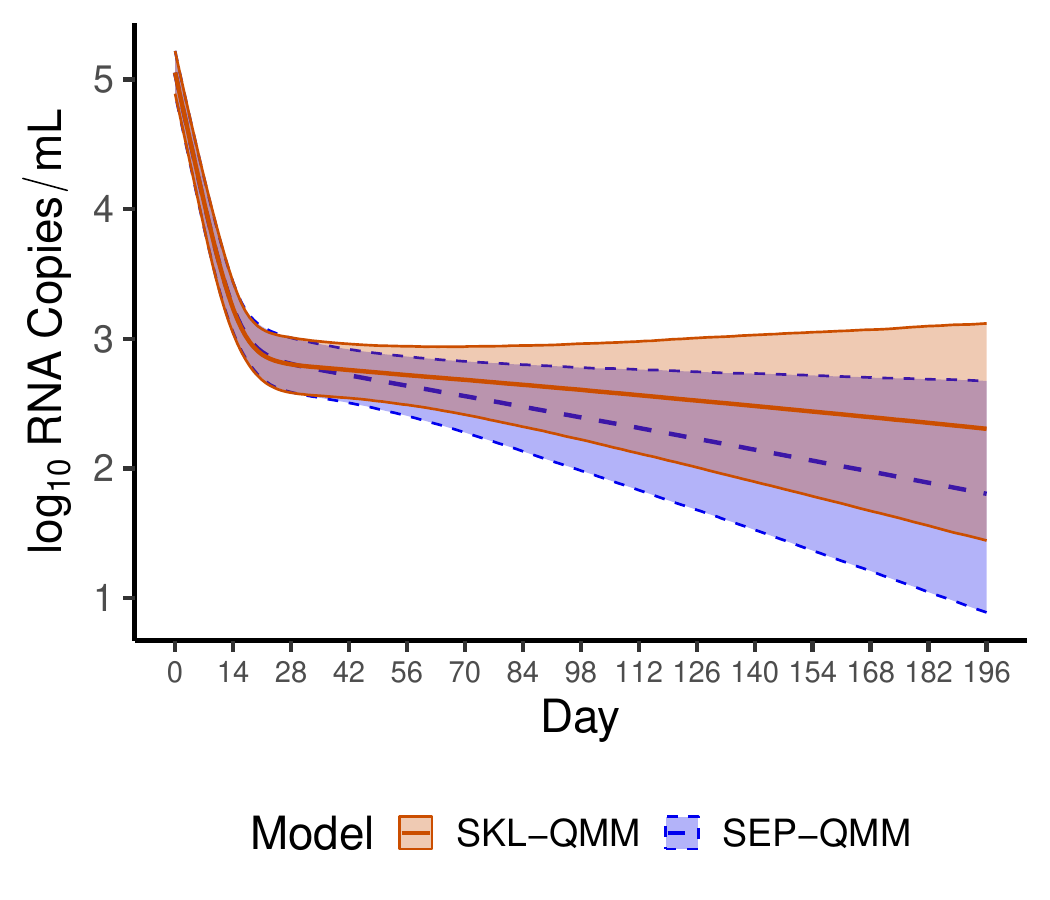}
            \caption{$p_0 = 0.50$}
        \end{subfigure}
        \begin{subfigure}{0.45\textwidth}
            \centering
            \includegraphics[width=\linewidth]{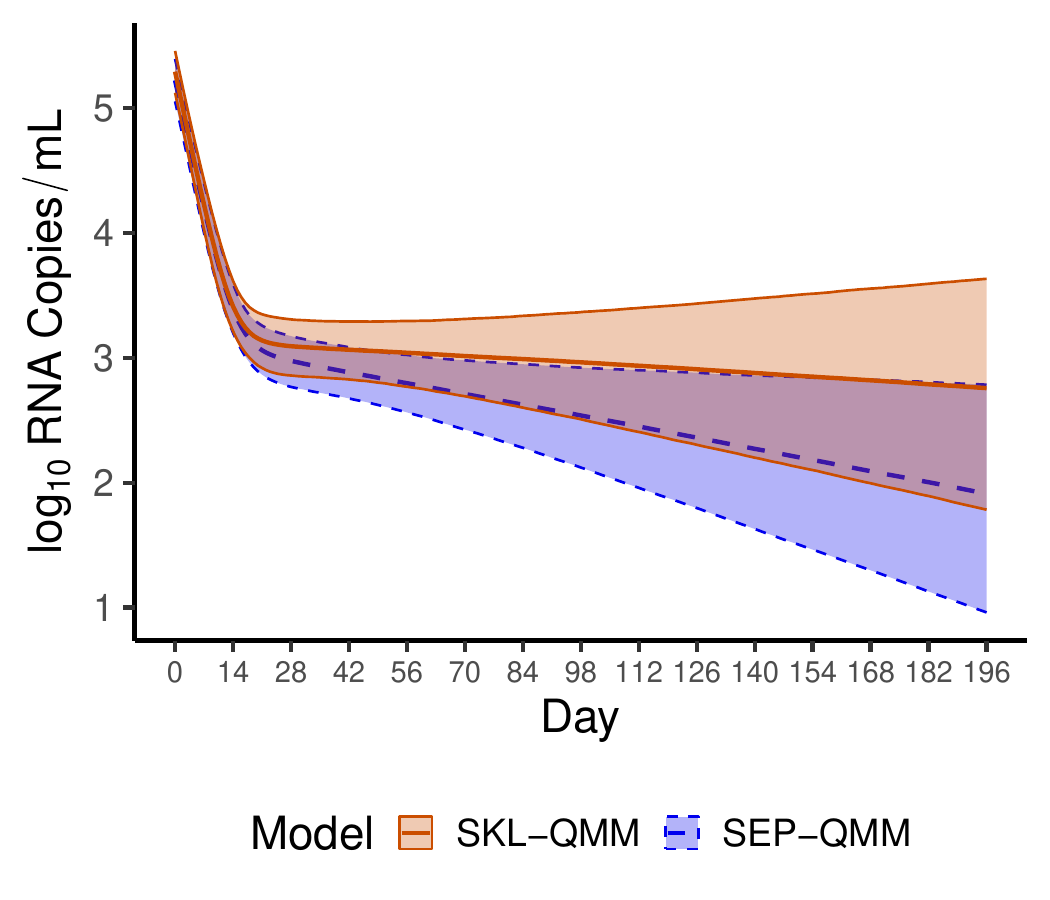}
            \caption{$p_0 = 0.75$}
        \end{subfigure}
        \\
        \begin{subfigure}{0.45\textwidth}
            \centering
            \includegraphics[width=\linewidth]{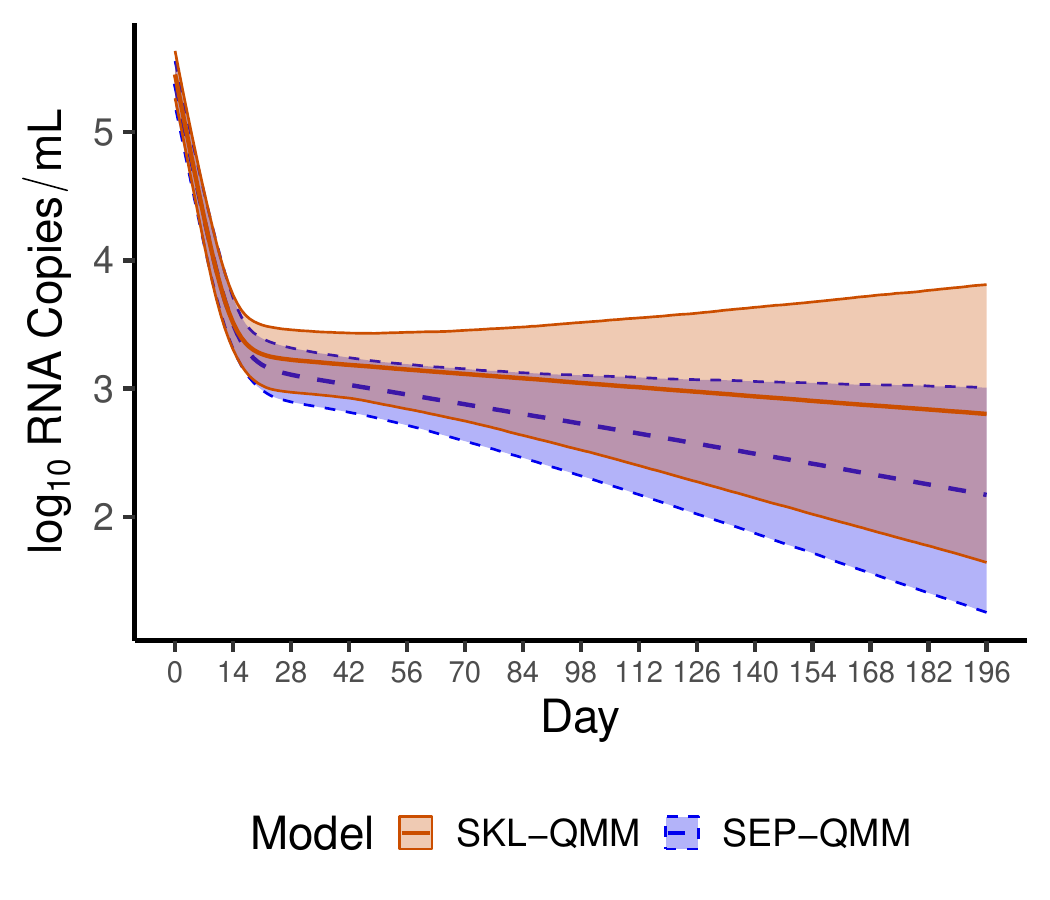}
            \caption{$p_0 = 0.90$}
        \end{subfigure}
        \caption{Quantile trajectories from the SL model (SKL-QMM; solid orange lines) and SEP model (SEP-QMM; dashed blue lines) for five quantiles ($p_0 = 0.10, 0.25, 0.50, 0.75, 0.90$). Shaded regions show 95\% BCIs around each trajectory, illustrating distinct decay patterns across lower, median, and upper viral load quantiles.}
    \label{fig:SEP_SKL_QUANTILE_PROFILES}
    \end{figure}

    \autoref{fig:SEP_SKL_QUANTILE_PROFILES} displays the population-level trajectories at five prespecified quantiles. All random effects are set to zero, so each curve represents a typical (population median) patient. We then fixed CD4 on a reference trajectory with a baseline of $2.25$ and a slope of $0.001$ (estimated from a simple linear mixed model). Across all five quantiles, the SL and SEP curves diverge visibly, most sharply in the late phase, showing that the different tail and skew specifications yield distinct population-level predictions.

    \afterpage{
        \begin{landscape}
            \thispagestyle{landscape}
            \begin{figure}
                \centering
                \includegraphics[width=1\linewidth, height=0.5\linewidth]{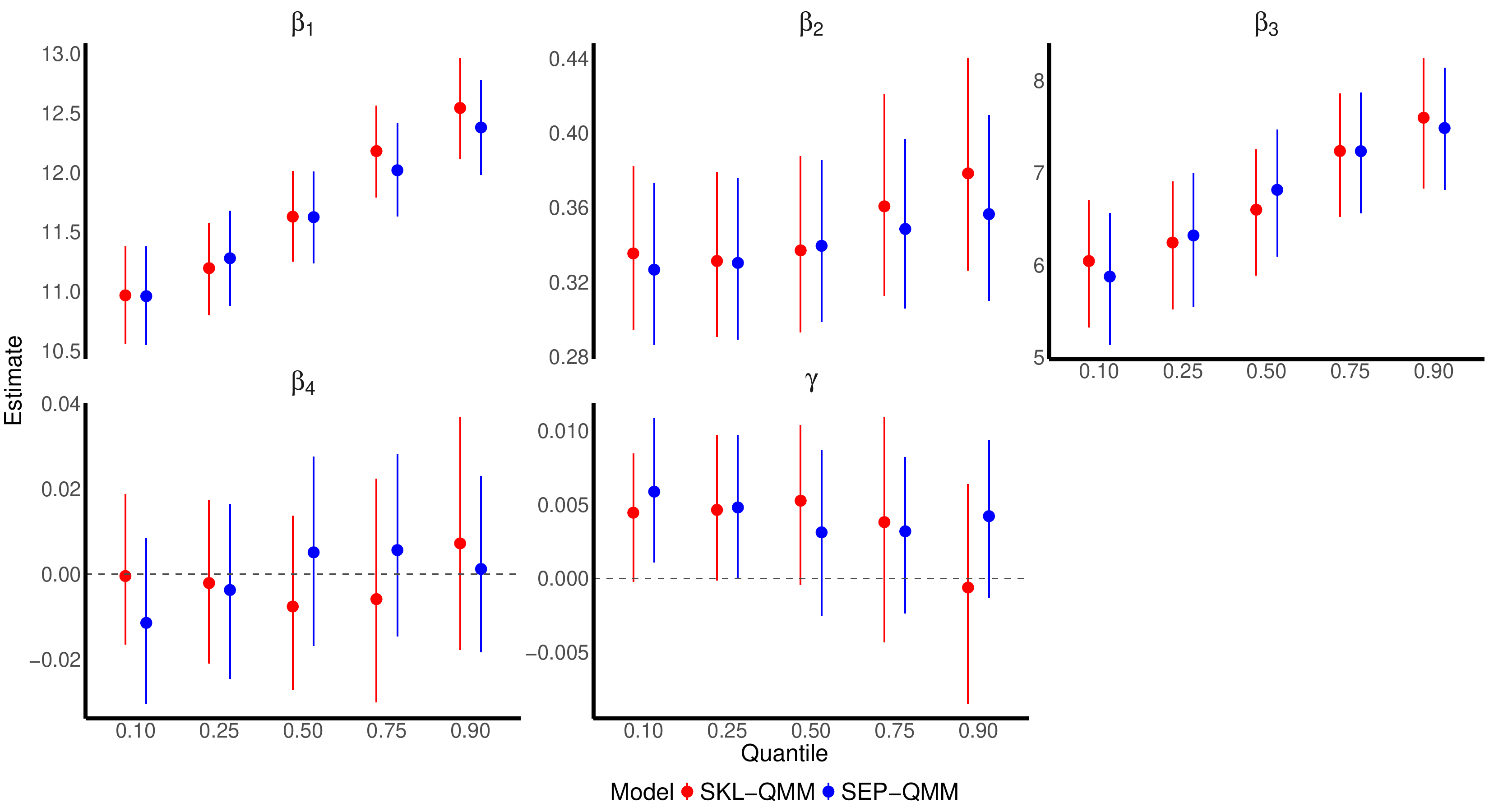}
                \caption{Posterior medians and 95\% BCIs for the biexponential decay parameters ($\beta_{1}, \beta_{2}, \beta_{3}, \beta_{4}$) and the CD4-influence parameter ($\gamma$) across quantiles $p_0 = 0.10, 0.25, 0.50, 0.75, 0.90$, comparing SKL-QMM (red) and SEP-QMM (blue). The dashed horizontal gray line indicates the zero value for the coefficient.}
                \label{fig:BETA_COMPARISON}
            \end{figure}
        \end{landscape}
    }
    
    \autoref{fig:BETA_COMPARISON} shows posterior medians and 95\% BCIs for the five fixed effect coefficients; full numeric summaries appear in Web \autoref{tab:PARAM_EST_CI} (Supplementary Material). Across quantiles, the SEP-QMM estimates $\kappa_{1}$ and $\kappa_{2}$ values that deviate markedly from the Laplace limit $\left(\kappa_{1}=\kappa_{2}=1\right)$, showing that tail weight, and therefore skewness, can adjust independently of the targeted quantile, which the SL benchmark cannot match.

    \subsection{Residual checks}

    Web \autoref{fig:QQ_OVERLAY} (Supplementary Material) overlays the \texttt{DHARMa} QQ plots for SKL-QMM and SEP-QMM for the five quantiles. Both models reproduce the observed residual distribution reasonably well, while the SEP curves track the 45$^{\circ}$ reference line more closely, especially at $p_{0}=0.75$ and $0.90$, implying a modest but consistent improvement in fit.

    \section{Simulation study} \label{sec:SIMULATION}

    The simulation study has two aims. First, when data are generated from the SEP distribution, we verify that the SEP-QMM recovers the true fixed effects and achieves nominal frequentist properties (minimal bias and 95\% coverage). Second, we investigate the impact of tail misspecification by analyzing the same datasets using the SKL-QMM, a special case obtained when $\kappa_{1} = \kappa_{2} = 1$. By sampling directly from SEP distributions with several left-right tail configurations and then fitting both the SEP-QMM and the SKL-QMM, we quantify how misspecified tails influence bias, root mean square error (RMSE), and coverage.

    We generated 300 synthetic datasets, each containing $N = 15$ subjects measured at five weekly visits (study days $0, 7, 14, 21, 28$). Centering the calendar at day 14 and dividing by seven to express time in weeks yields the scaled design points $t_{ij} = -2, -1, 0, 1, 2$. Subject-specific intercepts and slopes $\left(v_{i,1}, v_{i,2}\right)^{\top}$ were drawn from a bivariate normal distribution with mean $\left(0, 0\right)^{\top}$ and Cholesky factor
    \begin{equation}
        \mathbf{L}_v
        =\begin{pmatrix}
        3 & 0 \\
        0 & 1.5
        \end{pmatrix}.
    \end{equation}   
    Fixed effect targets were set to $\beta_{1} = 5$ and $\beta_{2} = -0.25$. The latent mean at each visit was $\mu_{ij} = \beta_{1} + v_{i,1} + \left(\beta_{2} + v_{i,2}\right)t_{ij}$. Conditional on these means, responses followed a SEP distribution $\operatorname{SEP}\left(\mu_{ij}, \sigma, \kappa_{1}, \kappa_{2}, p_{0}\right)$ with $\sigma = 0.40$ and three tail-parameter pairs $\left(2, 0.5\right)$, $\left(1, 1\right)$, and $\left(0.5, 2\right)$.

    Left censoring was imposed by replacing the smallest $c\%$ of generated responses with the corresponding $c\%$ empirical quantile, where $c$ took values 5\% and 10\%. This procedure yielded about 5\% and 11\% censored observations on average across replicates. Crossing the two censoring levels, two target quantiles $\left(p_{0} = 0.50, 0.80\right)$, and the three $\left(\kappa_{1},\kappa_{2}\right)$ settings produced 12 simulation scenarios.
    
    To avoid occasional sampler divergences in replicated datasets, we assigned independent $\kappa_{1},\kappa_{2}\sim\mathrm{Uniform}\left(0.01,3\right)$ in the simulation study; for the real data analysis, we retained the weakly informative $t_3^{+}\left(0,\sqrt{2}\right)$ prior used elsewhere in the paper.
    
    For $\beta_{1}$ and $\beta_{2}$, we recorded bias, RMSE, the average 95\% BCI interval length, and coverage probability (CP); the results are given in \autoref{tab:betas-skl-sep}. Across all scenarios, SEP-QMM stayed unbiased, and its 95\% BCI coverage remained near nominal. SKL-QMM diverged only when the heavy tail coincided with the part of the data that drives inference, namely, the median-quantile case with extreme tails and the $p_{0}=0.80$ scenario in which the uncensored right tail was heavy. When the heavy tail lies on the censored side at $p_{0}=0.80$ or when errors were truly Laplace, SKL-QMM performed like SEP-QMM. Hence, SEP-QMM's extra tail flexibility protects against misspecification without sacrificing efficiency when the simpler SL form is adequate.

    \afterpage{
    
        \begin{landscape}

            \thispagestyle{landscape}
        
            \begin{longtable}{
              S[table-format=1.2] 
              S[table-format=1.1] 
              S[table-format=1.1] 
              S[table-format=1.1] 
              l 
              S[table-format=2.2] 
              *{3}{S[table-format=2.4]}  S[table-format=1.3] 
              l 
              *{3}{S[table-format=2.4]}  S[table-format=1.3] 
            }
            
            \caption{
                Simulation study: performance of SKL-QMM and SEP-QMM in estimating the fixed effects $\beta_{1}$ and $\beta_{2}$ from data generated with SEP errors. Bias, RMSE, mean 95\% BCI length (Len), and CP are reported for three tail configurations $\left(\kappa_{1},\kappa_{2}\right)\in\left\{\left(2,0.5\right), \left(1,1\right), \left(0.5,2\right)\right\}$ and two censoring regimes ($\approx$5\% and 11\%). Results are stratified by the target quantile $p_{0}\in\left\{0.50,0.80\right\}$. The pairs $\left(2,0.5\right)$ and $\left(0.5,2\right)$ induce pronounced right and left skewness, respectively, whereas $\left(1,1\right)$ collapses the SEP error to the SL special case. Bias is the mean of the estimate minus the truth, Len is the average width of the 95\% BCIs, and CP is the proportion of BCIs that contain the truth.
            }
            \label{tab:betas-skl-sep}\\
            
            \toprule
             & & & & & & \multicolumn{9}{c}{\textbf{Model}} \\
             \cline{7-15}
             & & & & & &
             \multicolumn{4}{c}{\textbf{SKL-QMM}} & &
             \multicolumn{4}{c}{\textbf{SEP-QMM}}\\
            \cline{7-10}\cline{12-15}
            \textbf{Cen.} & $\bm{p_{0}}$ &
            $\bm{\kappa_{1}}$ & $\bm{\kappa_{2}}$ &
            \textbf{Param.} & \textbf{True} &
            \textbf{Bias} & \textbf{RMSE} & \textbf{Len} & \textbf{CP} & &
            \textbf{Bias} & \textbf{RMSE} & \textbf{Len} & \textbf{CP}\\
            \midrule
            \endfirsthead
            
            \multicolumn{15}{c}{\tablename\ \thetable\ -- \textit{Continued from previous page}}\\
            \toprule
             & & & & & & \multicolumn{9}{c}{\textbf{Model}} \\
             \cline{7-15}
             & & & & & &
             \multicolumn{4}{c}{\textbf{SKL-QMM}} & &
             \multicolumn{4}{c}{\textbf{SEP-QMM}}\\
            \cline{7-10}\cline{12-15}
            \textbf{Cen.} & $\bm{p_{0}}$ &
            $\bm{\kappa_{1}}$ & $\bm{\kappa_{2}}$ &
            \textbf{Param.} & \textbf{True} &
            \textbf{Bias} & \textbf{RMSE} & \textbf{Len} & \textbf{CP} & &
            \textbf{Bias} & \textbf{RMSE} & \textbf{Len} & \textbf{CP}\\
            \midrule
            \endhead
            
            \toprule
            \multicolumn{15}{r}{\textit{Continued on next page}}\\
            \endfoot
            
            \bottomrule
            \endlastfoot
            
            0.05 & 0.5 & 2.0 & 0.5 & $\beta_{1}$ & 5.00 & 0.1230 & 0.1610 & 0.4641 & 0.840 &  & 0.0254 & 0.1054 & 0.4430 & 0.963 \\* 
               &  &  &  & $\beta_{2}$ & -0.25 & 0.0031 & 0.1821 & 0.7487 & 0.933 &  & 0.0037 & 0.1808 & 0.7371 & 0.940 \\ 
              0.05 & 0.5 & 1.0 & 1.0 & $\beta_{1}$ & 5.00 & -0.0075 & 0.0939 & 0.4220 & 0.960 &  & -0.0077 & 0.0937 & 0.4315 & 0.967 \\* 
               &  &  &  & $\beta_{2}$ & -0.25 & 0.0124 & 0.1865 & 0.7301 & 0.953 &  & 0.0124 & 0.1863 & 0.7312 & 0.957 \\ 
              0.05 & 0.5 & 0.5 & 2.0 & $\beta_{1}$ & 5.00 & -0.1089 & 0.1514 & 0.4592 & 0.857 &  & -0.0110 & 0.1040 & 0.4425 & 0.973 \\* 
               &  &  &  & $\beta_{2}$ & -0.25 & 0.0149 & 0.1797 & 0.7525 & 0.957 &  & 0.0132 & 0.1786 & 0.7485 & 0.953 \\ 
              0.05 & 0.8 & 2.0 & 0.5 & $\beta_{1}$ & 5.00 & 0.0829 & 0.1400 & 0.4586 & 0.910 &  & 0.0252 & 0.1054 & 0.4344 & 0.957 \\* 
               &  &  &  & $\beta_{2}$ & -0.25 & 0.0056 & 0.1631 & 0.7247 & 0.963 &  & 0.0052 & 0.1618 & 0.7176 & 0.953 \\ 
              0.05 & 0.8 & 1.0 & 1.0 & $\beta_{1}$ & 5.00 & 0.0014 & 0.0971 & 0.4132 & 0.943 &  & 0.0150 & 0.0989 & 0.4199 & 0.943 \\* 
               &  &  &  & $\beta_{2}$ & -0.25 & 0.0136 & 0.1881 & 0.7340 & 0.927 &  & 0.0137 & 0.1879 & 0.7339 & 0.930 \\ 
              0.05 & 0.8 & 0.5 & 2.0 & $\beta_{1}$ & 5.00 & 0.0113 & 0.1025 & 0.4875 & 0.977 &  & 0.0055 & 0.0994 & 0.4650 & 0.983 \\* 
               &  &  &  & $\beta_{2}$ & -0.25 & -0.0061 & 0.1930 & 0.7459 & 0.960 &  & -0.0063 & 0.1911 & 0.7432 & 0.960 \\ 
              0.11 & 0.5 & 2.0 & 0.5 & $\beta_{1}$ & 5.00 & 0.1040 & 0.1486 & 0.4757 & 0.897 &  & 0.0120 & 0.1042 & 0.4543 & 0.970 \\* 
               &  &  &  & $\beta_{2}$ & -0.25 & -0.0155 & 0.1872 & 0.7650 & 0.937 &  & -0.0113 & 0.1848 & 0.7478 & 0.937 \\ 
              0.11 & 0.5 & 1.0 & 1.0 & $\beta_{1}$ & 5.00 & -0.0013 & 0.0917 & 0.4238 & 0.977 &  & -0.0007 & 0.0944 & 0.4322 & 0.980 \\* 
               &  &  &  & $\beta_{2}$ & -0.25 & 0.0089 & 0.1896 & 0.7357 & 0.933 &  & 0.0086 & 0.1894 & 0.7354 & 0.930 \\ 
              0.11 & 0.5 & 0.5 & 2.0 & $\beta_{1}$ & 5.00 & -0.1245 & 0.1602 & 0.4706 & 0.847 &  & -0.0296 & 0.1018 & 0.4504 & 0.963 \\* 
               &  &  &  & $\beta_{2}$ & -0.25 & 0.0024 & 0.1858 & 0.7352 & 0.937 &  & 0.0024 & 0.1831 & 0.7321 & 0.937 \\ 
              0.11 & 0.8 & 2.0 & 0.5 & $\beta_{1}$ & 5.00 & 0.0784 & 0.1302 & 0.4642 & 0.937 &  & 0.0294 & 0.1036 & 0.4435 & 0.963 \\* 
               &  &  &  & $\beta_{2}$ & -0.25 & -0.0106 & 0.1808 & 0.7327 & 0.930 &  & -0.0072 & 0.1802 & 0.7252 & 0.937 \\ 
              0.11 & 0.8 & 1.0 & 1.0 & $\beta_{1}$ & 5.00 & 0.0059 & 0.0918 & 0.4144 & 0.970 &  & 0.0185 & 0.0949 & 0.4211 & 0.960 \\* 
               &  &  &  & $\beta_{2}$ & -0.25 & -0.0087 & 0.1708 & 0.7259 & 0.933 &  & -0.0087 & 0.1708 & 0.7256 & 0.933 \\ 
              0.11 & 0.8 & 0.5 & 2.0 & $\beta_{1}$ & 5.00 & -0.0111 & 0.1009 & 0.4805 & 0.977 &  & -0.0110 & 0.0987 & 0.4626 & 0.973 \\* 
               &  &  &  & $\beta_{2}$ & -0.25 & 0.0111 & 0.1756 & 0.7615 & 0.957 &  & 0.0118 & 0.1735 & 0.7613 & 0.957 \\ 
            
            \end{longtable}
        
        \end{landscape}

    }
    
    \section{Discussion}\label{sec:DISCUSS}

    We have presented a Bayesian QMM that employs the SEP distribution for the error term. By giving the two shape parameters $\kappa_{1}$ and $\kappa_{2}$ control over tail weight and asymmetry, the SEP decouples skewness and kurtosis from the chosen quantile level $p_{0}$, a flexibility absent from the standard SL formulation.

    When applied to the ACTG~315 data, the SEP model achieved higher log marginal likelihoods at every quantile, with the largest gains at $p_{0}=0.75$ and $0.90$. Posterior estimates of $\kappa_{1}$ and $\kappa_{2}$ departed markedly from the Laplace limit $(1,1)$, confirming that the data require tail behavior that the SL model cannot represent.

    Across all simulation scenarios, the SEP-QMM remained unbiased, and its 95\% BCIs coverage was nominal. The SKL-QMM matched that accuracy, except when the quantile lies at the center with strongly non-Laplace tails; in those extreme-tail median cases, its bias increased, and coverage fell to approximately 85\%. Hence, the two extra shape parameters in SEP-QMM guard against tail misspecification while leaving efficiency unchanged when the simpler SL form is adequate.

    Although our illustrations used left-censored longitudinal biomarkers with a single random effects layer, the likelihood factorization is general. The same strategy extends to accelerated failure time models with right or interval censoring, to deeper multilevel hierarchies, and to joint longitudinal-survival models in which a SEP-based longitudinal submodel passes tail-flexible random effects to the survival component.
    
    Fully semiparametric quantile methods can, in principle, accommodate any tail behavior, but they often entail heavier computation and less interpretable parameters. The SEP retains a simple four-parameter kernel that spans a wide range, from lighter than normal to substantially heavier than Laplace, while its closed-form CDF keeps censored-likelihood contributions straightforward.

    Limitations remain. Our estimation relies on MCMC, which can become slow for very large cohorts; faster variational or Laplace-based approximations are a natural next step. In addition, we analyzed each quantile separately rather than borrowing strength across $p_{0}$, so the current approach does not enforce non-crossing of fitted quantile curves. Extending the model to jointly estimate multiple quantiles would address this, but it requires additional structure and is left for future work.

    In summary, the SEP-based QMM provides a robust and interpretable tool for datasets that combine hierarchical structure, censoring, skewness, and heavy tails. By allowing the data rather than the quantile index to dictate tail behavior, this approach provides a principled method for modeling the clinically informative extremes of the distribution.

    \section*{Acknowledgements}

    
    This work is based on the research supported by the National Research Foundation (NRF) of South Africa (Grant number 132383). Opinions expressed and conclusions arrived at are those of the authors and are not necessarily to be attributed to the NRF.
    
    \section*{Declaration of conflicting interests}

    The authors declare that they have no conflicts of interest.
    
    \section*{Data availability statement}

    All data utilized in this study are publicly accessible, and key parts of the application code, including the implementation of the proposed method, are available on \href{https://github.com/DABURGER1/Censored-SEP-Mixed-Model}{GitHub}.

    \begin{appendices}

        \section*{Appendix: Skew exponential power distribution CDF} \label{sec:SEP_CDF}

        In this appendix, we derive the CDF of the SEP distribution by splitting the calculation into two regions, $y \leq \mu$ and $y > \mu$.

        \Needspace{6\baselineskip}
        \noindent {\bf For} $\bm{y \leq \mu}$:
        
        When $y \leq \mu$, the PDF takes the form:
        \begin{equation}
            f_{\text{SEP}}\left(t\right) = \frac{1}{\sigma}
            \exp\left(-\frac{1}{\kappa_1}\left(\frac{\mu - t}{2 p_0 \sigma K_1}\right)^{\kappa_1}\right).
        \end{equation}
        We start from:
        \begin{equation}
            F_{\text{SEP}}\left(y\right) = \int_{-\infty}^{y} f_{\text{SEP}}\left(t\right) \mathrm{d}t.
        \end{equation}
        The integration limits are $-\infty$ to $y \leq \mu$.
        
        Make the substitution:
        \begin{equation}
            u = \frac{1}{\kappa_1}\left(\frac{\mu - t}{2 p_0 \sigma K_1}\right)^{\kappa_1}.
        \end{equation}
        When $t=y\leq\mu$, we have $\mu-y\geq0$ and thus $u_y\geq0$. As $t\to -\infty$, we have $\mu - t \to \infty$, so $u \to \infty$.
        
        This change of variables reverses the direction of integration. Initially, the integral is from $t=-\infty$ to $t=y$. In terms of $u$, when $t=-\infty$, $u=\infty$, and when $t=y$, $u=u_y$. Thus:
        \begin{equation}
            \int_{-\infty}^{y} \cdots \mathrm{d}t = \int_{\infty}^{u_y} \cdots \mathrm{d}u.
        \end{equation}
        Rewriting $\mu - t$:
        \begin{equation}
            \mu - t = 2 p_0 \sigma K_1 \left(\kappa_1 u\right)^{1/\kappa_1}.
        \end{equation}
        Differentiating with respect to $t$:
        \begin{equation}
            \mathrm{d}t = -\frac{2 p_0 \sigma K_1 \left(\kappa_1\right)^{1/\kappa_1}}{\kappa_1} u^{\frac{1}{\kappa_1}-1} \mathrm{d}u.
        \end{equation}
        Using the definition of $K_1$:
        \begin{equation}
            K_1 = \frac{\left(\kappa_1\right)^{-1/\kappa_1}}{2\Gamma\left(1+\frac{1}{\kappa_1}\right)}
            \implies 2 p_0 \sigma K_1 \left(\kappa_1\right)^{1/\kappa_1} = \frac{p_0 \sigma}{\Gamma\left(1+\frac{1}{\kappa_1}\right)}.
        \end{equation}
        Thus:
        \begin{equation}
            \mathrm{d}t = -\frac{p_0 \sigma}{\kappa_1 \Gamma\left(1+\frac{1}{\kappa_1}\right)} u^{\frac{1}{\kappa_1}-1}\mathrm{d}u.
        \end{equation}
        Substitute into $F_{\text{SEP}}\left(y\right)$:
        \begin{align}
            F_{\text{SEP}}\left(y\right) &= \int_{-\infty}^{y} f_{\text{SEP}}\left(t\right)\mathrm{d}t \\
            &= \int_{\infty}^{u_y} \frac{1}{\sigma}\exp\left(-u\right)\left(-\frac{p_0 \sigma}{\kappa_1\Gamma\left(1+\frac{1}{\kappa_1}\right)}u^{\frac{1}{\kappa_1}-1}\right)\mathrm{d}u.
        \end{align}
        Pulling out constants and flipping the integration limits to restore the conventional order (from lower to higher):
        \begin{equation}
            F_{\text{SEP}}\left(y\right) = p_0 \int_{u_y}^{\infty} \frac{u^{\frac{1}{\kappa_1}-1}\exp\left(-u\right)}{\Gamma\left(\frac{1}{\kappa_1}\right)} \mathrm{d}u,
        \end{equation}
        since $\kappa_1\Gamma\left(1+\frac{1}{\kappa_1}\right) = \Gamma\left(\frac{1}{\kappa_1}\right)$.
        
        Notice that this integral now runs from $u_y$ to $\infty$. The regularized incomplete gamma function $G\left(a,b\right)$ is defined as:
        \begin{equation}
            G\left(a,b\right) = \frac{1}{\Gamma\left(b\right)}\int_0^a t^{b-1}\exp\left(-t\right)\mathrm{d}t.
        \end{equation}
        For $b=\frac{1}{\kappa_1}$, the integral from $0$ to $\infty$ of this kernel is 1. Therefore:
        \begin{equation}
            \int_{0}^{\infty} \frac{u^{\frac{1}{\kappa_1}-1}\exp\left(-u\right)}{\Gamma\left(\frac{1}{\kappa_1}\right)}\mathrm{d}u = 1.
        \end{equation}
        Thus:
        \begin{equation}
            \int_{u_y}^{\infty} \frac{u^{\frac{1}{\kappa_1}-1}\exp\left(-u\right)}{\Gamma\left(\frac{1}{\kappa_1}\right)}\mathrm{d}u 
            = 1 - \int_{0}^{u_y} \frac{u^{\frac{1}{\kappa_1}-1}\exp\left(-u\right)}{\Gamma\left(\frac{1}{\kappa_1}\right)}\mathrm{d}u 
            = 1 - G\left(u_y,\tfrac{1}{\kappa_1}\right),
        \end{equation}
        since the integral from $0$ to $u_y$ defines $G\left(u_y,\tfrac{1}{\kappa_1}\right)$.
        
        Substitute this back into $F_{\text{SEP}}\left(y\right)$:
        \begin{equation}
            F_{\text{SEP}}\left(y\right) = p_0 \left[1 - G\left(u_y,\frac{1}{\kappa_1}\right)\right].
        \end{equation}
        Finally, substitute $u_y$:
        \begin{equation}
            F_{\text{SEP}}\left(y\right) = p_0\left[1 - G\left(\frac{1}{\kappa_1}\left(\frac{\mu - y}{2 p_0 \sigma K_1}\right)^{\kappa_1}, \frac{1}{\kappa_1}\right)\right] \quad \text{for } y \leq \mu.
        \end{equation}
        This completes the derivation for $y \leq \mu$.
    
        \Needspace{6\baselineskip}
        {\bf For} $\bm{y > \mu}$:
        
        For $y > \mu$, an analogous approach is applied. The PDF becomes:
        \begin{equation}
            f_{\text{SEP}}\left(t\right) = \frac{1}{\sigma}\exp\left(-\frac{1}{\kappa_2}\left(\frac{t-\mu}{2\left(1 - p_0\right)\sigma K_2}\right)^{\kappa_2}\right).
        \end{equation}
        We write:
        \begin{equation}
            F_{\text{SEP}}\left(y\right) = \int_{-\infty}^{\mu} f_{\text{SEP}}\left(t\right)\mathrm{d}t + \int_{\mu}^{y} f_{\text{SEP}}\left(t\right)\mathrm{d}t.
        \end{equation}
        By construction, the first integral equals $p_0$. The second integral, after a similar substitution for $u$ and recognizing the incomplete gamma form, yields:
        \begin{equation}
            F_{\text{SEP}}\left(y\right) = p_0 + \left(1 - p_0\right) G\left(\frac{1}{\kappa_2}\left(\frac{y-\mu}{2\left(1 - p_0\right)\sigma K_2}\right)^{\kappa_2}, \frac{1}{\kappa_2}\right) \quad \text{for } y > \mu.
        \end{equation}

    \end{appendices}

    \renewcommand{\refname}{References}
    \input{bbl.bbl}

    \newpage
    \clearpage

    \markboth{}{}

    \setcounter{page}{1}
    \setcounter{figure}{0}
    \setcounter{table}{0}
    \renewcommand{\restoreapp}{}
    \renewcommand\appendixname{Web Appendix}
    \renewcommand{\figurename}{Web Figure}
    \renewcommand{\tablename}{Web Table}
    \renewcommand{\headrulewidth}{0pt}
    \titleformat{\section}{\large\bfseries}{\appendixname~\thesection .}{0.5em}{}
    
    \begin{appendices}
    
        {\Large\bf A flexible quantile mixed-effects model for censored outcomes}
        
        \vskip 1.0cm
        
        {\normalsize Divan A. Burger, Sean van der Merwe, and Emmanuel Lesaffre}
        
        \vskip 4.5truecm
        
        \begin{center}
            \noindent
            {\Large\bf Supporting Information}
        \end{center}

        \newpage
        \clearpage

        \begin{landscape}


        \thispagestyle{landscape}

        \singlespacing
        
        \begin{longtable}{l l S[table-format=3.3] S[table-format=4.3] S[table-format=3.3] l S[table-format=3.3] S[table-format=4.3] S[table-format=3.3]}
            \caption{Posterior summaries for each quantile level under the SKL-QMM and SEP-QMM models. Columns report each parameter's posterior estimate (Est.) and 95\% BCI. Results for higher quantiles continue on subsequent pages.} \label{tab:PARAM_EST_CI}\\
            \toprule
            & & \multicolumn{7}{c}{{\bf Model}} \\
            \cline{3-9}
            & & \multicolumn{3}{c}{{\bf SKL-QMM}} & & \multicolumn{3}{c}{{\bf SEP-QMM}} \\
            \cline{3-5} \cline{7-9}
            \textbf{Quantile} & 
            \textbf{Parameter} & 
            \textbf{Est.} & 
            \multicolumn{2}{c}{{\bf 95\% BCI}} &
            \textbf{} &
            \textbf{Est.} & 
            \multicolumn{2}{c}{{\bf 95\% BCI}} \\
            \midrule
            \endfirsthead
            
            \multicolumn{9}{l}{\textit{(Continued from previous page)}}\\
            \toprule
            & & \multicolumn{7}{c}{{\bf Model}} \\
            \cline{3-9}
            & & \multicolumn{3}{c}{{\bf SKL-QMM}} & & \multicolumn{3}{c}{{\bf SEP-QMM}} \\
            \cline{3-5} \cline{7-9}
            \textbf{Quantile} & 
            \textbf{Parameter} & 
            \textbf{Est.} & 
            \multicolumn{2}{c}{{\bf 95\% BCI}} &
            \textbf{} &
            \textbf{Est.} & 
            \multicolumn{2}{c}{{\bf 95\% BCI}} \\
            \midrule
            \endhead
            
            \toprule
            \multicolumn{9}{r}{\textit{(Continued on next page)}}\\
            \endfoot
            
            \bottomrule
            \endlastfoot
            
            0.10 & $\beta_{1}$ & 10.967 & [10.557; & 11.378] &  & 10.960 & [10.550; & 11.382] \\* 
               & $\beta_{2}$ & 0.336 & [0.294; & 0.382] &  & 0.327 & [0.286; & 0.374] \\* 
               & $\beta_{3}$ & 6.044 & [5.323; & 6.704] &  & 5.875 & [5.134; & 6.565] \\* 
               & $\beta_{4}$ & -0.000 & [-0.017; & 0.019] &  & -0.011 & [-0.030; & 0.008] \\* 
               & $\gamma$ & 0.004 & [-0.000; & 0.008] &  & 0.006 & [0.001; & 0.011] \\* 
               & $\sigma$ & 0.079 & [0.068; & 0.091] &  & 0.516 & [0.301; & 0.735] \\* 
               & $L_{v,11}$ & 1.081 & [0.790; & 1.883] &  & 1.261 & [0.864; & 2.889] \\* 
               & $L_{v,21}$ & 3.088 & [-3.491; & 19.925] &  & 6.472 & [-2.549; & 42.260] \\* 
               & $L_{v,31}$ & -0.496 & [-1.264; & -0.106] &  & -0.708 & [-2.346; & -0.239] \\* 
               & $L_{v,41}$ & 18.852 & [-8.147; & 42.576] &  & 24.555 & [-6.257; & 53.384] \\* 
               & $L_{v,22}$ & 12.081 & [8.187; & 20.017] &  & 13.748 & [8.912; & 25.259] \\* 
               & $L_{v,32}$ & 0.042 & [-0.361; & 0.511] &  & -0.175 & [-0.728; & 0.302] \\* 
               & $L_{v,42}$ & -21.377 & [-48.438; & -1.123] &  & -12.589 & [-42.280; & 8.150] \\* 
               & $L_{v,33}$ & 0.687 & [0.516; & 0.902] &  & 0.724 & [0.541; & 0.972] \\* 
               & $L_{v,43}$ & -24.646 & [-39.780; & -9.773] &  & -25.945 & [-39.794; & -12.913] \\* 
               & $L_{v,44}$ & 31.635 & [22.530; & 45.085] &  & 28.466 & [20.433; & 39.604] \\* 
               & $\kappa_{1}$ &  &  &  &  & 0.353 & [0.249; & 0.521] \\* 
               & $\kappa_{2}$ &  &  &  &  & 1.431 & [0.857; & 2.418] \\ 
              0.25 & $\beta_{1}$ & 11.196 & [10.801; & 11.579] &  & 11.280 & [10.881; & 11.678] \\* 
               & $\beta_{2}$ & 0.331 & [0.291; & 0.379] &  & 0.330 & [0.289; & 0.376] \\* 
               & $\beta_{3}$ & 6.244 & [5.519; & 6.907] &  & 6.321 & [5.550; & 6.998] \\* 
               & $\beta_{4}$ & -0.002 & [-0.021; & 0.017] &  & -0.004 & [-0.025; & 0.017] \\* 
               & $\gamma$ & 0.005 & [-0.000; & 0.010] &  & 0.005 & [0.000; & 0.010] \\* 
               & $\sigma$ & 0.179 & [0.156; & 0.207] &  & 0.334 & [0.175; & 0.549] \\* 
               & $L_{v,11}$ & 1.325 & [0.862; & 3.382] &  & 1.304 & [0.858; & 3.543] \\* 
               & $L_{v,21}$ & 9.591 & [-2.317; & 50.713] &  & 10.335 & [-3.214; & 54.953] \\* 
               & $L_{v,31}$ & -0.697 & [-2.526; & -0.191] &  & -0.733 & [-2.848; & -0.200] \\* 
               & $L_{v,41}$ & 13.374 & [-29.028; & 42.451] &  & 11.941 & [-32.942; & 42.834] \\* 
               & $L_{v,22}$ & 15.401 & [9.486; & 28.384] &  & 16.665 & [10.083; & 31.147] \\* 
               & $L_{v,32}$ & -0.092 & [-0.639; & 0.378] &  & -0.125 & [-0.767; & 0.437] \\* 
               & $L_{v,42}$ & -26.722 & [-57.830; & -3.347] &  & -27.887 & [-59.983; & -2.821] \\* 
               & $L_{v,33}$ & 0.702 & [0.528; & 0.932] &  & 0.738 & [0.552; & 1.007] \\* 
               & $L_{v,43}$ & -25.110 & [-40.125; & -10.698] &  & -25.965 & [-40.495; & -11.838] \\* 
               & $L_{v,44}$ & 30.592 & [22.065; & 43.045] &  & 29.560 & [21.486; & 40.696] \\* 
               & $\kappa_{1}$ &  &  &  &  & 0.569 & [0.396; & 0.881] \\* 
               & $\kappa_{2}$ &  &  &  &  & 0.833 & [0.573; & 1.323] \\ 
              0.50 & $\beta_{1}$ & 11.630 & [11.250; & 12.014] &  & 11.626 & [11.237; & 12.011] \\* 
               & $\beta_{2}$ & 0.337 & [0.293; & 0.388] &  & 0.340 & [0.299; & 0.386] \\* 
               & $\beta_{3}$ & 6.601 & [5.887; & 7.255] &  & 6.816 & [6.090; & 7.470] \\* 
               & $\beta_{4}$ & -0.008 & [-0.027; & 0.014] &  & 0.005 & [-0.017; & 0.028] \\* 
               & $\gamma$ & 0.005 & [-0.000; & 0.010] &  & 0.003 & [-0.003; & 0.009] \\* 
               & $\sigma$ & 0.252 & [0.219; & 0.291] &  & 0.326 & [0.170; & 0.519] \\* 
               & $L_{v,11}$ & 1.701 & [0.959; & 4.619] &  & 1.393 & [0.888; & 4.061] \\* 
               & $L_{v,21}$ & 16.257 & [-0.481; & 56.769] &  & 11.702 & [-2.974; & 55.485] \\* 
               & $L_{v,31}$ & -1.033 & [-3.355; & -0.290] &  & -0.832 & [-3.227; & -0.254] \\* 
               & $L_{v,41}$ & 9.083 & [-37.311; & 45.708] &  & 4.484 & [-43.176; & 37.299] \\* 
               & $L_{v,22}$ & 15.686 & [9.713; & 27.606] &  & 16.734 & [10.441; & 30.021] \\* 
               & $L_{v,32}$ & -0.161 & [-0.761; & 0.406] &  & -0.106 & [-0.729; & 0.523] \\* 
               & $L_{v,42}$ & -25.824 & [-54.943; & -2.215] &  & -32.226 & [-63.995; & -8.878] \\* 
               & $L_{v,33}$ & 0.745 & [0.553; & 1.036] &  & 0.763 & [0.562; & 1.048] \\* 
               & $L_{v,43}$ & -26.275 & [-40.439; & -12.445] &  & -24.042 & [-38.728; & -9.760] \\* 
               & $L_{v,44}$ & 29.182 & [20.984; & 40.393] &  & 29.582 & [21.585; & 41.071] \\* 
               & $\kappa_{1}$ &  &  &  &  & 0.910 & [0.573; & 1.542] \\* 
               & $\kappa_{2}$ &  &  &  &  & 0.619 & [0.453; & 0.838] \\ 
              0.75 & $\beta_{1}$ & 12.181 & [11.790; & 12.564] &  & 12.020 & [11.629; & 12.416] \\* 
               & $\beta_{2}$ & 0.361 & [0.313; & 0.421] &  & 0.349 & [0.306; & 0.397] \\* 
               & $\beta_{3}$ & 7.237 & [6.522; & 7.861] &  & 7.235 & [6.562; & 7.873] \\* 
               & $\beta_{4}$ & -0.006 & [-0.030; & 0.022] &  & 0.006 & [-0.015; & 0.028] \\* 
               & $\gamma$ & 0.004 & [-0.004; & 0.011] &  & 0.003 & [-0.002; & 0.008] \\* 
               & $\sigma$ & 0.195 & [0.168; & 0.227] &  & 0.485 & [0.294; & 0.688] \\* 
               & $L_{v,11}$ & 1.436 & [0.904; & 4.449] &  & 1.548 & [0.933; & 3.830] \\* 
               & $L_{v,21}$ & 7.550 & [-6.938; & 49.243] &  & 14.171 & [-2.345; & 50.258] \\* 
               & $L_{v,31}$ & -0.861 & [-3.738; & -0.175] &  & -0.998 & [-3.142; & -0.315] \\* 
               & $L_{v,41}$ & 14.660 & [-21.708; & 43.619] &  & -0.212 & [-47.487; & 34.962] \\* 
               & $L_{v,22}$ & 12.883 & [7.788; & 29.901] &  & 16.414 & [10.242; & 29.083] \\* 
               & $L_{v,32}$ & -0.149 & [-0.829; & 0.768] &  & -0.114 & [-0.791; & 0.587] \\* 
               & $L_{v,42}$ & -15.744 & [-43.382; & 4.056] &  & -34.414 & [-66.343; & -11.086] \\* 
               & $L_{v,33}$ & 0.744 & [0.541; & 1.006] &  & 0.783 & [0.581; & 1.066] \\* 
               & $L_{v,43}$ & -19.175 & [-31.978; & -6.277] &  & -23.613 & [-38.105; & -9.698] \\* 
               & $L_{v,44}$ & 26.968 & [18.816; & 38.890] &  & 29.518 & [21.684; & 40.325] \\* 
               & $\kappa_{1}$ &  &  &  &  & 1.693 & [0.980; & 3.186] \\* 
               & $\kappa_{2}$ &  &  &  &  & 0.486 & [0.376; & 0.624] \\ 
              0.90 & $\beta_{1}$ & 12.544 & [12.113; & 12.967] &  & 12.380 & [11.979; & 12.783] \\* 
               & $\beta_{2}$ & 0.378 & [0.326; & 0.440] &  & 0.357 & [0.310; & 0.410] \\* 
               & $\beta_{3}$ & 7.598 & [6.830; & 8.249] &  & 7.486 & [6.817; & 8.143] \\* 
               & $\beta_{4}$ & 0.007 & [-0.018; & 0.037] &  & 0.001 & [-0.018; & 0.023] \\* 
               & $\gamma$ & -0.001 & [-0.009; & 0.006] &  & 0.004 & [-0.001; & 0.009] \\* 
               & $\sigma$ & 0.084 & [0.071; & 0.099] &  & 0.676 & [0.480; & 0.855] \\* 
               & $L_{v,11}$ & 1.029 & [0.770; & 1.680] &  & 1.810 & [0.955; & 4.957] \\* 
               & $L_{v,21}$ & -0.647 & [-18.125; & 16.707] &  & 19.816 & [1.009; & 59.327] \\* 
               & $L_{v,31}$ & -0.522 & [-1.492; & -0.111] &  & -1.260 & [-4.178; & -0.378] \\* 
               & $L_{v,41}$ & 16.787 & [1.232; & 36.328] &  & -6.102 & [-53.580; & 29.292] \\* 
               & $L_{v,22}$ & 10.117 & [6.583; & 31.363] &  & 15.751 & [9.924; & 27.693] \\* 
               & $L_{v,32}$ & -0.069 & [-0.707; & 0.784] &  & -0.241 & [-0.895; & 0.238] \\* 
               & $L_{v,42}$ & -7.638 & [-29.739; & 8.835] &  & -32.189 & [-61.950; & -10.163] \\* 
               & $L_{v,33}$ & 0.715 & [0.531; & 0.936] &  & 0.735 & [0.553; & 0.976] \\* 
               & $L_{v,43}$ & -16.996 & [-27.568; & -6.061] &  & -23.761 & [-37.640; & -10.907] \\* 
               & $L_{v,44}$ & 23.644 & [16.704; & 33.666] &  & 28.525 & [21.089; & 38.630] \\* 
               & $\kappa_{1}$ &  &  &  &  & 2.979 & [1.646; & 6.455] \\* 
               & $\kappa_{2}$ &  &  &  &  & 0.349 & [0.278; & 0.437] \\ 

        \end{longtable}
        
        \end{landscape}

        \newpage

        \begin{figure}[!b]
            \centering
            \begin{subfigure}{0.44\textwidth}
                \centering
                \includegraphics[width=\linewidth]{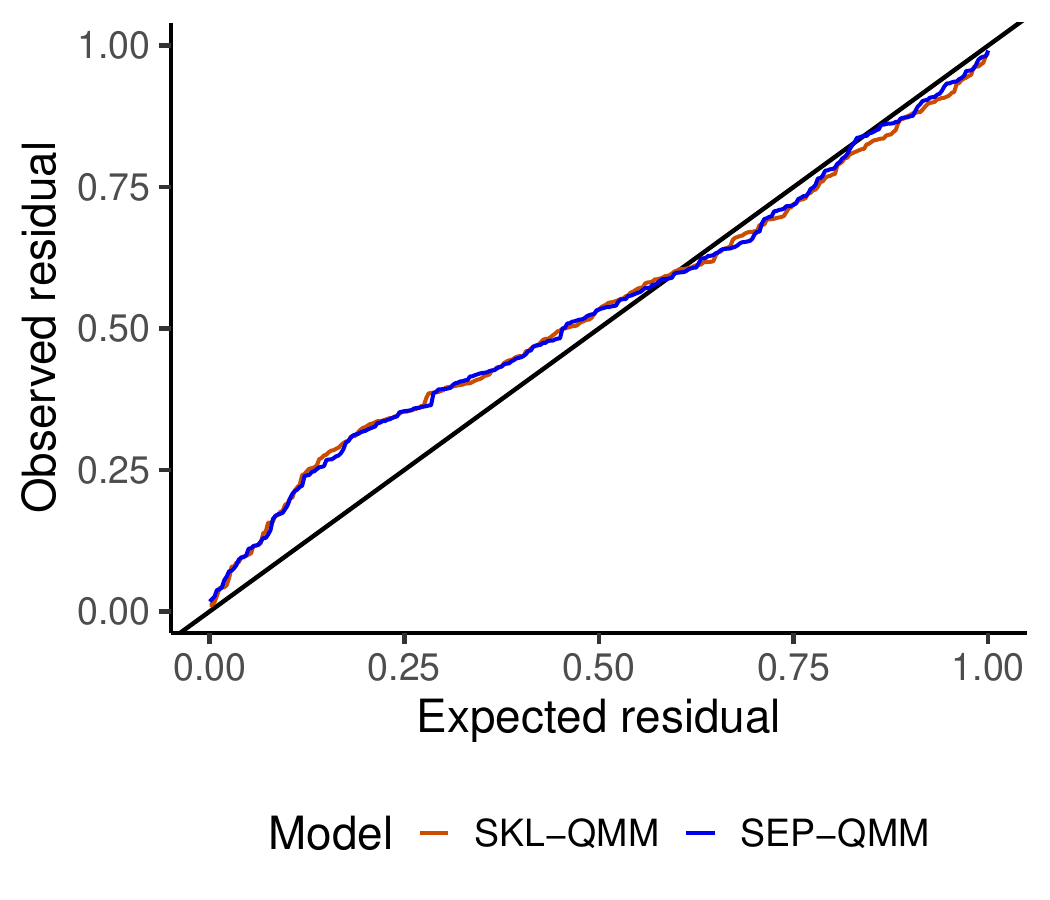}
                \caption{$p_0 = 0.10$}
            \end{subfigure}
            \begin{subfigure}{0.44\textwidth}
                \centering
                \includegraphics[width=\linewidth]{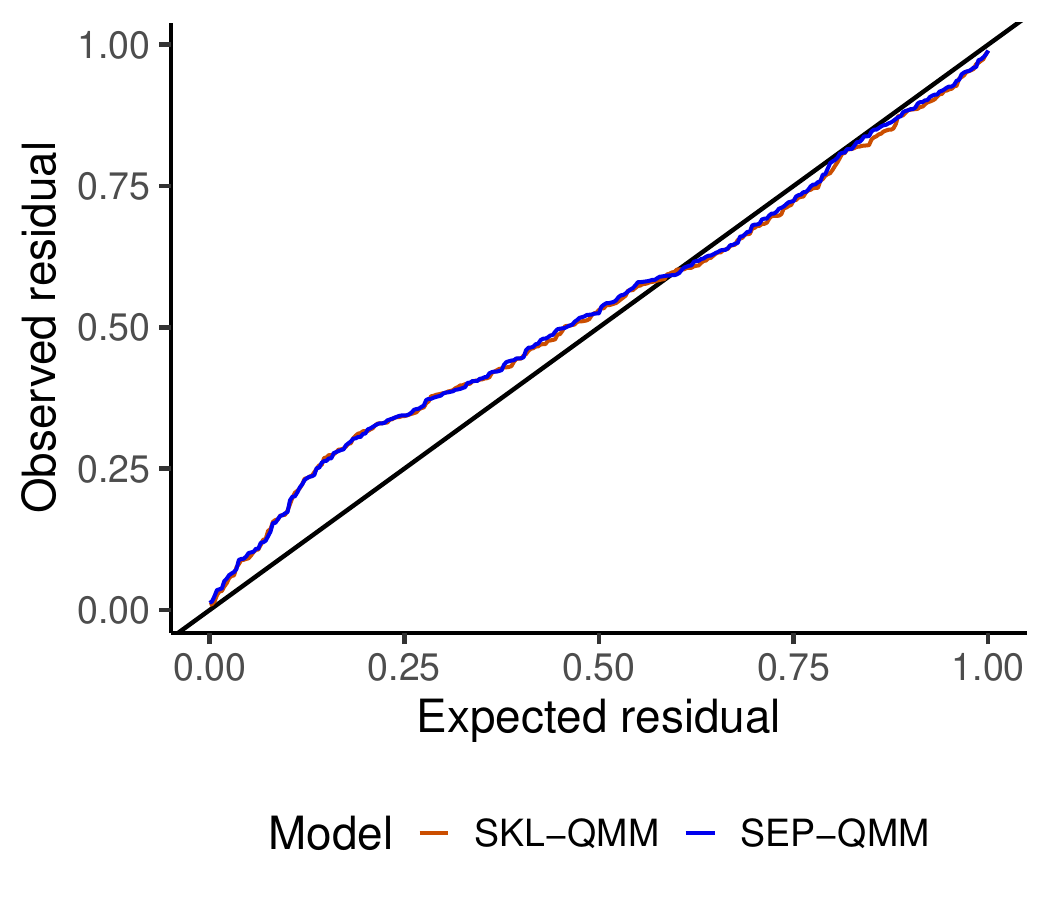}
                \caption{$p_0 = 0.25$}
            \end{subfigure}
            \\
            \begin{subfigure}{0.44\textwidth}
                \centering
                \includegraphics[width=\linewidth]{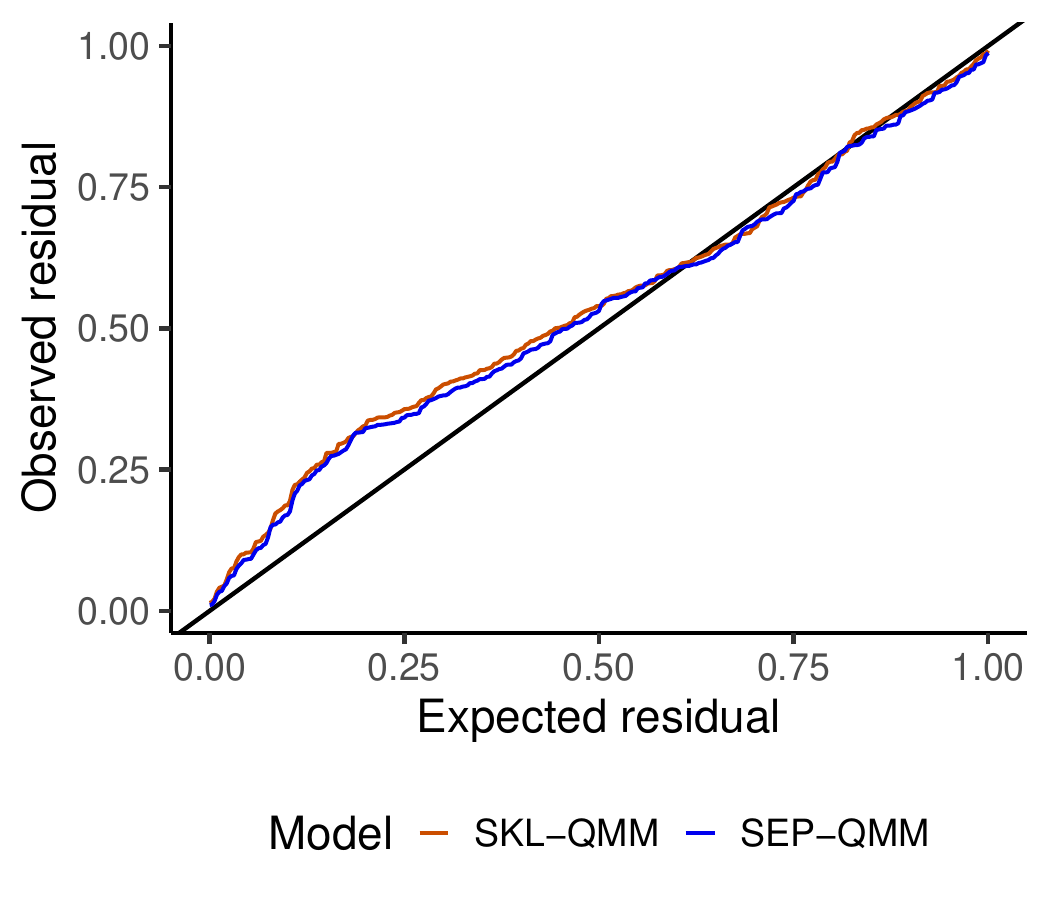}
                \caption{$p_0 = 0.50$}
            \end{subfigure}
            \begin{subfigure}{0.44\textwidth}
                \centering
                \includegraphics[width=\linewidth]{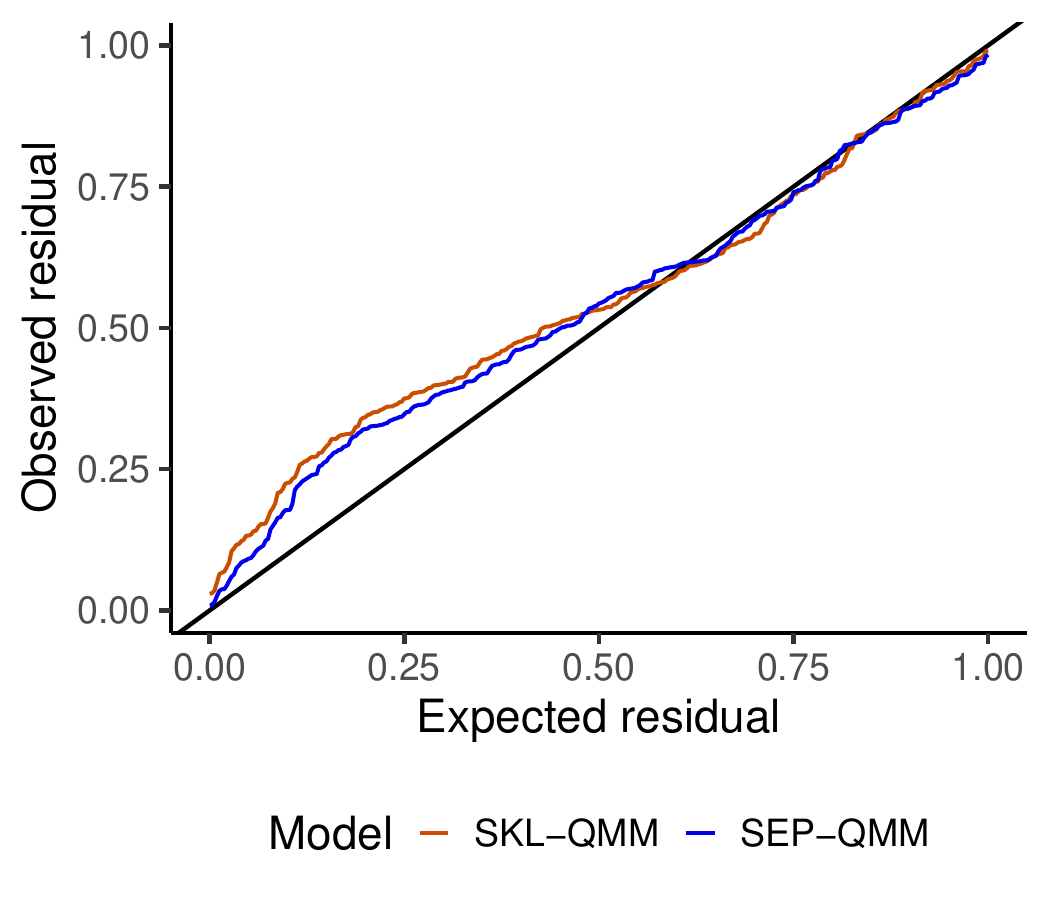}
                \caption{$p_0 = 0.75$}
            \end{subfigure}
            \\
            \begin{subfigure}{0.44\textwidth}
                \centering
                \includegraphics[width=\linewidth]{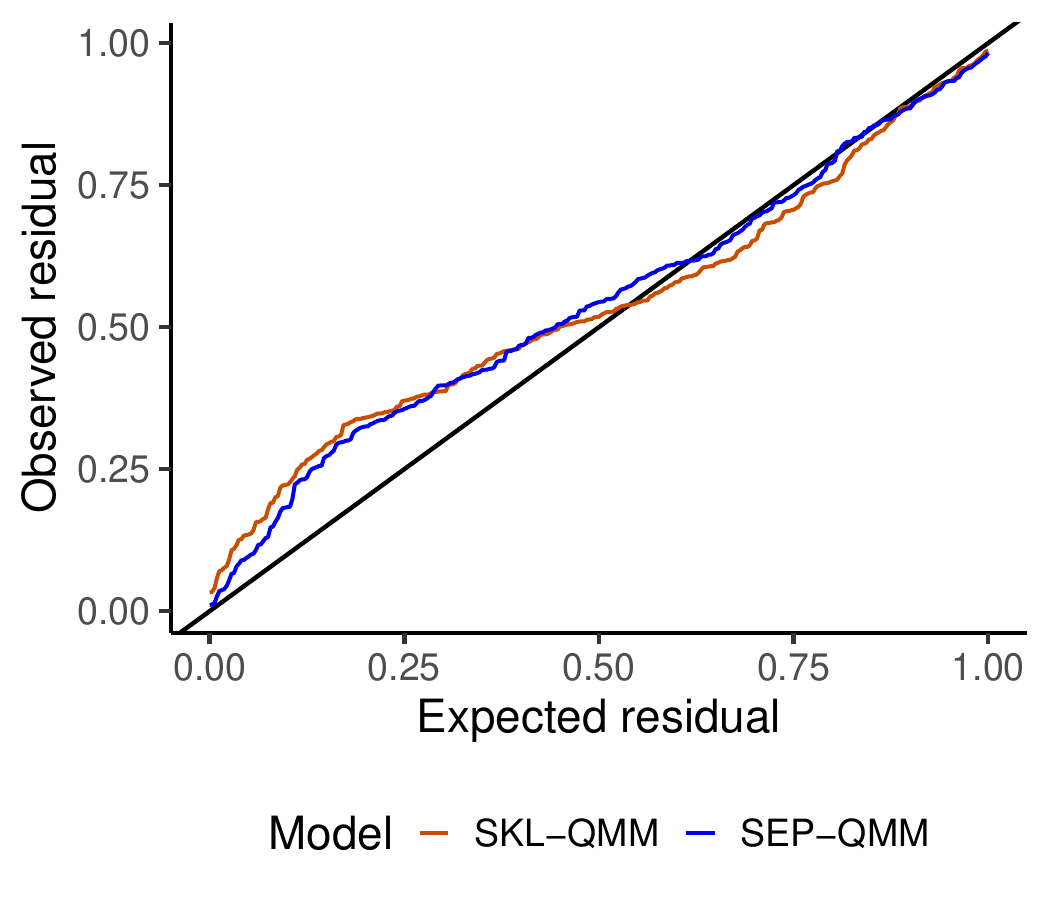}
                \caption{$p_0 = 0.90$}
            \end{subfigure}
            \caption{Overlay QQ plots comparing posterior predictive residuals from the two error structures. The red and blue curves represent SKL-QMM and SEP-QMM, respectively, at $p_0 = 0.10$, $0.25$, $0.50$, $0.75$, and $0.90$. The thin $45^{\circ}$ diagonal marks perfect calibration; curves that lie closer to this line indicate the more adequate model.} \label{fig:QQ_OVERLAY}
        \end{figure}
        
    \end{appendices}
    
\end{document}